\documentclass{aastex631}

\usepackage{amsmath}
\usepackage{booktabs}

\usepackage{chngpage}
\usepackage{multirow}
\usepackage{subfigure}
\usepackage{threeparttable} 
\usepackage{float}
\usepackage{hyperref}
\usepackage[section]{placeins} 
\usepackage[figuresright]{rotating}

\usepackage{color}
\shorttitle{The Oscillation in Evolution of Changing-look Blazar OQ 334}
\shortauthors{Ren et al.}

\begin{document}

\title{The Oscillation in Evolution of Changing-look Blazar OQ 334 }

\correspondingauthor{Y. G. Zheng }
\email{$^\ast$ynzyg@ynu.edu.cn}
\email{$^\dagger$kangshiju@alumni.hust.edu.cn}

\author[0009-0003-0317-4831]{S. S. Ren}
\affiliation{Department of Physics, Yunnan Normal University, Kunming, Yunnan, 650092, People's Republic of China}

\author{R. X. Zhou}
\affiliation{Shandong Key Laboratory of Optical Astronomy and Solar-Terrestrial Environment, School of Space Science and Physics, Institute of Space Sciences, Shandong University, Weihai, Shandong, 264209, China}

\author[0000-0003-0170-9065]{Y. G. Zheng$^\ast$}
\affiliation{Department of Physics, Yunnan Normal University, Kunming, Yunnan, 650092, People's Republic of China}

\author[0000-0002-9071-5469]{S. J. Kang$^\dagger$}
\affiliation{School of Physics and Electrical Engineering, Liupanshui Normal University, Liupanshui, Guizhou, 553004, People's Republic of China}

\begin{abstract}

%We investigate the evolution of changing-look blazar (CLB) on long time scales as to expect to trace the state change of CLB. Three morphological types including FSRQ state, transition state, and BL Lac state are classified according to the criteria proposed by analyzing the relationship between the equivalent width (EW) of the emission line and the $\gamma$-ray photon spectral index $\Gamma_{\gamma}$. The multi-wavelength light curves and spectral energy distributions (SEDs) corresponding to different epochs are obtained. The efforts found that $\Gamma_{\gamma}$ satisfy the relationshipes with $\Gamma_{\gamma} $ {\bf\color{blue}$\gtrsim$} 2.2 for FSRQ state,  $2.0 < \Gamma_{\gamma}  <  2.2$ for transition state, and $\Gamma_{\gamma}$ {\bf\color{blue}$\lesssim$} 2.0 for BL Lac state. We apply the criteria to the photon spectrum evolution of CLB OQ 334 during MJD 58678 - 60387. The evolution is subdivided into 5 FSRQ states, 9 transition states, and 4 BL Lac states. Moreover, we use the model spectra parameters of each state epoch to test the reliability of subdivided morphological types. The result shows that : (1) the accretion rate parameter is consistent with our earlier research; (2)  there is an increasing trend in the epochs of the BL Lac states, even if there is not an obvious decreasing trend in epochs of the FSRQ states. We issue that strong evidence that CLB is a especial epoch in the evolution of blazars could be obtained from the oscillation phenomenon in the CLB evolution.

We investigate the evolution of a changing-look blazar (CLB) on long timescales and expect to trace the state change of a CLB. Three morphological types, including a flat spectrum radio quasar (FSRQ) state, transition state, and BL Lacertae (BL Lac) state are classified according to the criteria proposed by analyzing the relationship between the equivalent width of the emission line and the $\gamma$-ray photon spectral index $\Gamma_{\gamma}$. The multiwavelength light curves and spectral energy distributions corresponding to different epochs are obtained. The efforts found that $\Gamma_{\gamma}$ satisfy the relationships with $\Gamma_{\gamma} $ $\gtrsim$ 2.2 for the FSRQ state,  $2.0 < \Gamma_{\gamma}  <  2.2$ for the transition state, and $\Gamma_{\gamma}$ $\lesssim$ 2.0 for the BL Lac state. We apply the criteria to the photon spectrum evolution of CLB OQ 334 during MJD 58678 - 60387. The evolution is subdivided into five FSRQ states, nine transition states, and four BL Lac states. Moreover, we use the model spectra parameters of each state epoch to test the reliability of subdivided morphological types. The result shows that: (1) the accretion rate parameter is consistent with our earlier research; and (2)  there is an increasing trend in the epochs of the BL Lac states, even if there is not an obvious decreasing trend in epochs of the FSRQ states. We issue that strong evidence that a CLB is an especial epoch in the evolution of blazars that could be obtained from the oscillation phenomenon in the CLB evolution.

\end{abstract}


\section{Introduction} \label{sec:intro}
Blazars are a unique subclass of active galactic nuclei, distinguished by their emission line characteristics and high-energy emissions.
These objects have two primary categories (\citealt{1991ApJS...76..813S,1995PASP..107..803U}): BL Lacertae (BL Lac) objects, which exhibit very weak or no emission lines with an equivalent width (EW) of less than 5 {\AA}, and flat spectrum radio quasars (FSRQs), which have a strong emission line with an EW greater than 5 {\AA}.

FSRQs and BL Lacs possess a spectral energy distribution (SED) that spans the entire electromagnetic spectrum.
Typically, the SED of a blazar features two prominent humps.
Researchers attributed the low-energy hump to synchrotron (Syn) radiation (\citealt{1998AdSpR..21...89U}).
Meanwhile, the origin of the high-energy hump remains a topic of debate.
%, with two main radiation models proposed: the leptonic model and the hadronic model.

In the leptonic model (\citealt{1992ApJ...397L...5M}), the high-energy hump is produced either by the synchrotron self-Compton (SSC; \citealt{1985ApJ...298..114M}) process, where the seed photons originate from Syn radiation within the jet, or by the external Compton (EC) process, where the seed photons come from external sources such as the accretion disk (\citealt{1992A&A...256L..27D,1994ApJ...421..153S}), broad-line region (BLR; \citealt{1994ApJ...421..153S,1995ApJ...441...79B,1996MNRAS.280...67G,1998MNRAS.294..439B}), dust torus, or the cosmic microwave background radiation (\citealt{1995ARA&A..33..163W}).

Blazars can also be categorized based on the peak frequency of their Syn radiation ($v_{peak}^{syn}$) into three types (\citealt{2010ApJ...716...30A}): low synchrotron-peaked blazars (LSP; $v_{peak}^{syn} < 10^{14}$ Hz), intermediate synchrotron-peaked blazars ($10^{14}$ Hz $< v_{peak}^{syn} < 10^{15}$ Hz), and high synchrotron-peaked blazars ($v_{peak}^{syn} > 10^{15} $Hz). Generally, FSRQs are then a subclass of LSPs, together with low-frequency-peaked BL Lac (LBLs).
Correspondingly, we categorize LBL ($v_{peak,BLLac}^{syn} < 10^{14}$ Hz), intermediate-frequency-peaked BL Lac ($10^{14}$ Hz $< v_{peak,BLLac}^{syn} < 10^{15}$ Hz), and high-frequency-peaked BL Lac ($v_{peak,BLLac}^{syn} > 10^{15} $Hz). 
%In contrast, the peak frequency of Syn radiation in FSRQs is generally less than $10^{14}$ Hz.

With ongoing research, a unique class of active blazars emerged that alternates between the FSRQ and BL Lac state over several years (\citealt{2013MNRAS.432L..66G,2014ApJ...797...19R,2021ApJ...913..146M}).
These objects that refer to the change of the EW of the emission lines are defined as changing-look blazars (CLBs). 
The timescale for these blazars to switch from the FSRQ state to the BL Lac state appears to be getting shorter, while the duration in the BL Lac state is becoming longer (\citealt{2024A&A...685A.140R}).
This fact suggests that CLBs may represent a specific phase in the evolutionary process of blazars (\citealt{2022ApJ...927..227L,2023MNRAS.525.3201K}).

One such blazar, OQ 334, quiescent until 2018, was characterized as an FSRQ (\citealt{2020ApJS..247...33A}).
However, following a series of flares in 2018, significant changes in its broad emission lines were observed.
For instance, the EW of Mg II line measured $34_{-0.97}^{+0.97}$ on MJD 53472 then dropped to $4.5_{-2.1}^{+2.1}$ on MJD 58121.5 before increasing again (\citealt{2021ApJ...913..146M}). 
These variations in the EW of Mg II indicate that OQ 334 is a CLB (\citealt{2021ApJ...913..146M}).
It exhibited a BL Lac feature in January 2018 that lasted for 2 days, followed by an FSRQ emission line feature from April 2018 to June 2019 for 416 days, and then reverted to a BL Lac feature in 2019 July, which lasted 12 days (\citealt{2021ApJ...913..146M}).
Over time, OQ 334 has spent shorter epochs in the FSRQ state and increasingly longer epochs in the BL Lac state.
These facts raise the question: if OQ 334 continues to spend progressively longer epochs in the BL Lac state, will it eventually become a BL Lac object? To explore this we build on our previous work (\citealt{2024A&A...685A.140R}).

With this objective, we investigate the evolution of OQ 334 from the perspective of Fermi-Large Area Telescope (LAT) and aim to discern its evolutionary trajectory. We organized the paper as follows: Section 2 presents the data collection process. Section 3 outlines our selection criteria and model framework. Section 4 details the results obtained from our fitting analysis. Section 5 provides a comprehensive discussion. Finally, Section 6 summarizes our conclusions. Throughout our analysis, we adhere to the cosmological constants: $H_{0}=73.3$ km s$^{-1}$ Mpc$^{-1}$, $\Omega_{\rm M}=0.308$, and $\Omega_{\rm r}=0$, $\Omega_{\Lambda}=0.692$ (\citealt{2022ApJ...934L...7R}).%2016A&A...594A..13P

\section{Data Reduction and Analysis}   
\label{sec:data}
\subsection{$\gamma$-ray and \textit{X-Ray} Data}
We processed Fermi-LAT $\gamma$-ray light curves and data points in MJD 54628-60387 using the method outlined by \cite{2024A&A...685A.140R}.
X-ray data points were derived from Swift observations spanning MJD 58678-60387. 
Notably, these X-ray data were consolidated from multiple IDs (refer to column (6) of Table~\ref{tab:1}) into single events. For instance, during epoch $T_{4}$, which had 11 and 12 observation IDs, we following the standard threads\footnote{\scriptsize{\url{ https://www.swift.ac.uk/analysis/xrt/index.php}} in Swift-XRT and first} using the \textit{xselect} and \textit{ximage} components of the \textit{HEASOFT} software to combine these events and images, followed by \textit{xrtmkarf} to generate the response file (\citealt{2024ApJ...962...22Z}).
Finally, we utilized \textit{grppha} to derive the average X-ray spectrum for the ${T_4}$ epoch.
Similar methods were sequentially applied to compute X-ray data for other epochs (${T_4}$, ${T_5}$, ${B_3}$, ${T_6}$, ${F_4}$, ${F_5}$, ${F_6}$, ${T_{11}}$, ${B_6}$, ${T_{12}}$, and ${F_7}$).
Due to lower count numbers in some observations, we employed the Cash statistics method (\citealt{1979ApJ...228..939C}) for analyzing ungrouped data. 
For observations with higher count rates, we ensured a minimum of 20-30 counts per bin and used chi-square minimization for fitting. X-ray data for epochs without observation IDs (${F_3}$, ${T_7}$, ${B_4}$, ${T_8}$, ${T_9}$, ${B_5}$, and ${T_{10}}$) were aggregated from all observations in MJD 58678-60387.

\subsection{Optical/Ultraviolet and Other data}
Following the UVOT data analysis thread\footnote{\scriptsize{\url{ https://www.swift.ac.uk/analysis/uvot/index.php}}}, we merged images acquired with the same filter and Swift observation IDs using the \textit{uvotimsum} tool before running \textit{uvotsource}. Subsequently, we processed and redden corrected UVOT images summed across six filters for epochs ${T_4}$, ${T_5}$, ${B_3}$, ${T_6}$, ${F_4}$, ${F_5}$, ${F_6}$, ${T_{11}}$, ${B_6}$, ${T_{12}}$, and ${F_7}$ (refer to column (6) of Table~\ref{tab:1}).
Ultraviolet/optical data postprocessed from all observation IDs in MJD 58678-60387 use as data for epochs ${F_3}$, ${T_7}$, ${B_4}$, ${T_8}$, ${T_9}$, ${B_5}$, and ${T_{10}}$.

\subsection{g-band Data}   
All-Sky Automated Survey for Supernovae (ASAS-SN; \citealt{2014ApJ...788...48S,2017PASP..129j4502K}) is an automated program designed to routinely measure bright transient sources across the visible sky with minimal observational deviations, at a depth of about 17 mag. ASAS-SN initially consisted of two platforms, one at the Cerro Tololo International Observatory (CTIO) in Chile and the other at the Haleakala Observatory in Hawaii. In 2017, a second observatory was added to CTIO, along with one each at the McDonald Observatory in Texas and the South African Astrophysical Observatory.

ASAS-SN can observe both the $g$ and $V$ bands. 
\cite{2023ApJ...950..152A} indicated that the central wavelengths of the $g$ band and $V$ band are $\sim$480 nm and $\sim$551 nm, respectively.
The $V$ band is the observation of the two original stations. The $g$ band is the result of observations from three new stations in 2017. To limit the peak frequency of synchrotron radiation, we obtained the $g$-band observations of OQ 334 in the MJD 58678-60387 from ASAS-SN Sky Patrol\footnote{\scriptsize{\url{ https://asas-sn.osu.edu/}}}. For the corresponding $g$-band data for each epoch (see column (7) of Table~\ref{tab:1}), the arithmetic average is used.

\section{The model}

%------------------ jet power -----------------------------------
This work utilizes the models and methodologies detailed in \cite{2024A&A...685A.140R} to analyze the wavelength SEDs across different epochs.

We assume that the jet power is carried by relativistic electrons, protons, the magnetic field, and radiation. All of these components can be formally expressed as
%-------------------------------------------------------------------
\begin{equation}
P_{\mathrm{i}}=\pi R^2 \Gamma^2 c u_{\mathrm{i}}^{\prime}
\label{eq:1}
\end{equation}
%-------------------------------------------------------------------
where $u_{\mathrm{i}}^{\prime}$ denotes the comoving energy density of the $i$ component, which are given by
%-------------------------------------------------------------------
\begin{equation}
u_{\mathrm{e}}^{\prime}=m_e c^2 \int \gamma N(\gamma) d \gamma
\label{eq:2}
\end{equation}
%-------------------------------------------------------------------

%-------------------------------------------------------------------
\begin{equation}
u_{\mathrm{p}}^{\prime}=m_{\mathrm{p}} c^2 \int N(\gamma) d \gamma
\label{eq:3}
\end{equation}
%-------------------------------------------------------------------
%-------------------------------------------------------------------
\begin{equation}
u_{\mathrm{B}}^{\prime}=B^2 / 8 \pi
\label{eq:4}
\end{equation}
%-------------------------------------------------------------------

%-------------------------------------------------------------------
\begin{equation}
u_{\mathrm{rad}}^{\prime}=\frac{L_{o b s}}{4 \pi R^2 c \delta^4} \simeq \frac{L}{4 \pi R^2 c \delta^4}
\label{eq:5}
\end{equation}
%-------------------------------------------------------------------
where $L_{obs}$ denotes the total observed nonthermal luminosity and $L$ is the nonthermal luminosity derived from the modeling. $\gamma$ is the Lorentz factor. $N(\gamma)$ denotes the particle number density.

\section{Result} 
%---------------------------Fermi-LAT----------------------------------------
\subsection{\textit{Fermi}-LAT Light Curve and SED}
%-------------------------------------------------------------------
\begin{figure*}
\centering
\includegraphics[width=0.85\textwidth]{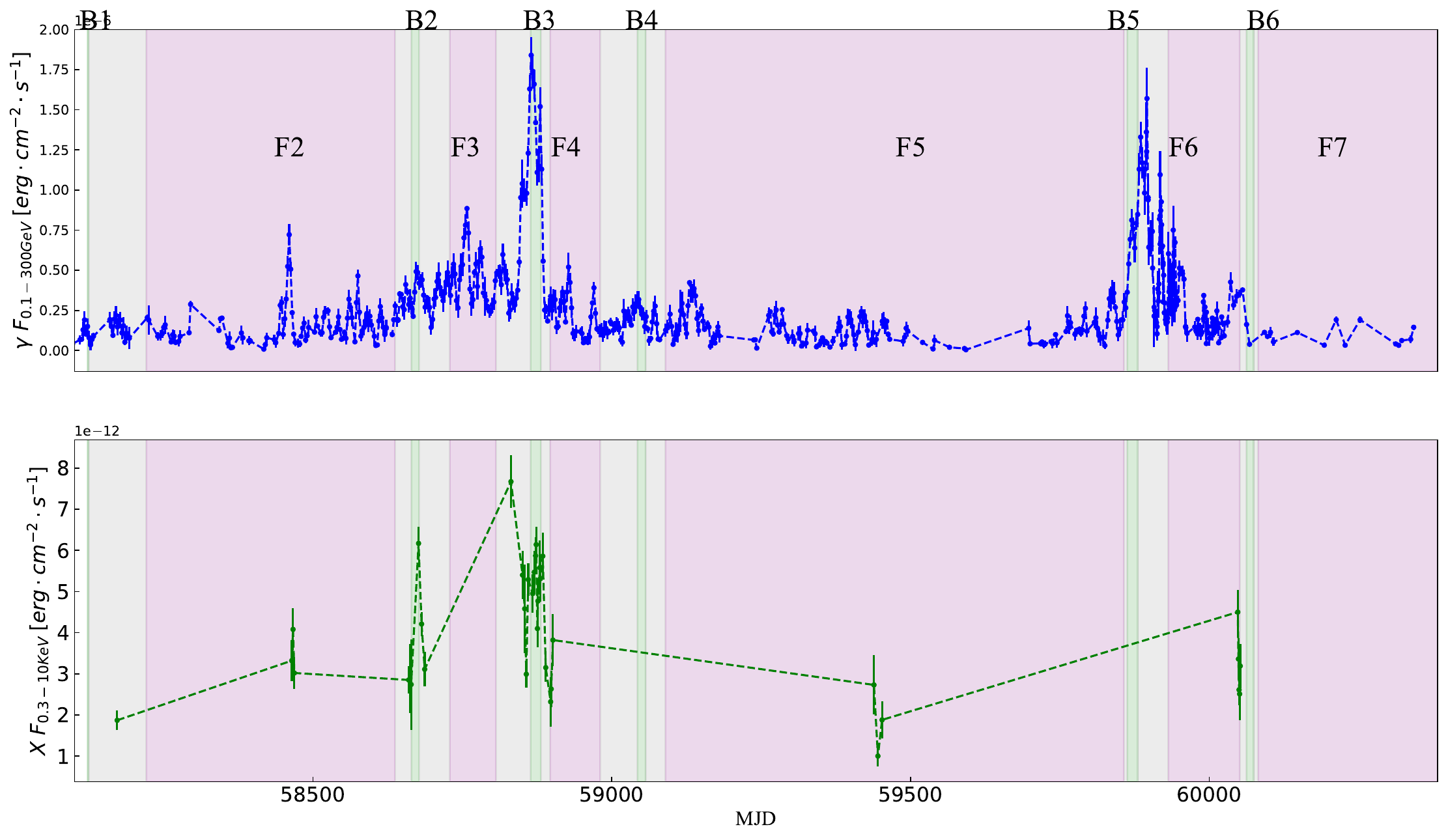}
\caption{$\gamma$-ray and X-ray light curve in MJD 54628-60387. The upper panel shows $\gamma$-ray data ranging from 0.1 to 300 GeV, while the lower panel displays X-ray data spanning 0.3-10 keV. Labels such as ${F_1}$ and ${B_1}$  indicate the first occurrence of OQ 334 in the FSRQ and BL Lac states, respectively. We discarded data points with $TS \geq 15$. The corresponding MJD for the epochs B1, F2, and B2 are from \cite{2024A&A...685A.140R}.}
\label{Fig01}
\end{figure*}
%-------------------------------------------------------------------

%%-------------------------------------------------------------------
\begin{figure}
	\centering
		\includegraphics[width=0.8\linewidth]{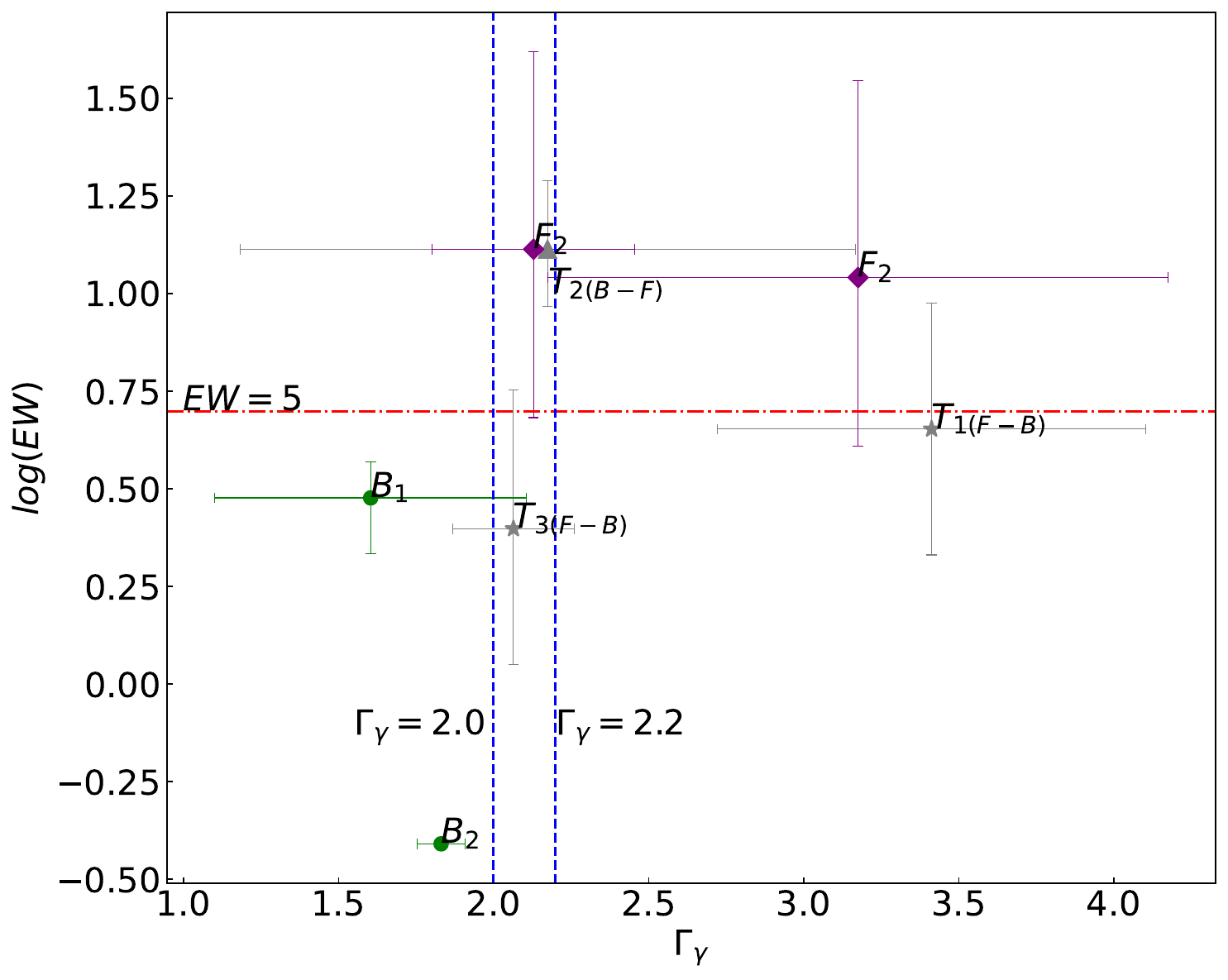}%EW_1
  \centering
\caption{Comparison of $\gamma$-ray photon spectral index from \cite{2024A&A...685A.140R} with the equivalent width of Mg \uppercase\expandafter{\romannumeral2} from \cite{2021ApJ...913..146M}. Green triangles, purple prisms, and gray stars represent the BL Lac state, FSRQ state, and transition state, respectively. The dashed red line represents $EW = 5 {\AA}$, while the two dotted blue lines correspond to = 2.0 and = 2.2} %
	\label{Fig_index_EW}
\end{figure}
%-------------------------------------------------------------------

%-------------------------------------------------------------------
\begin{figure*}
	\centering
		\includegraphics[width=1\linewidth]{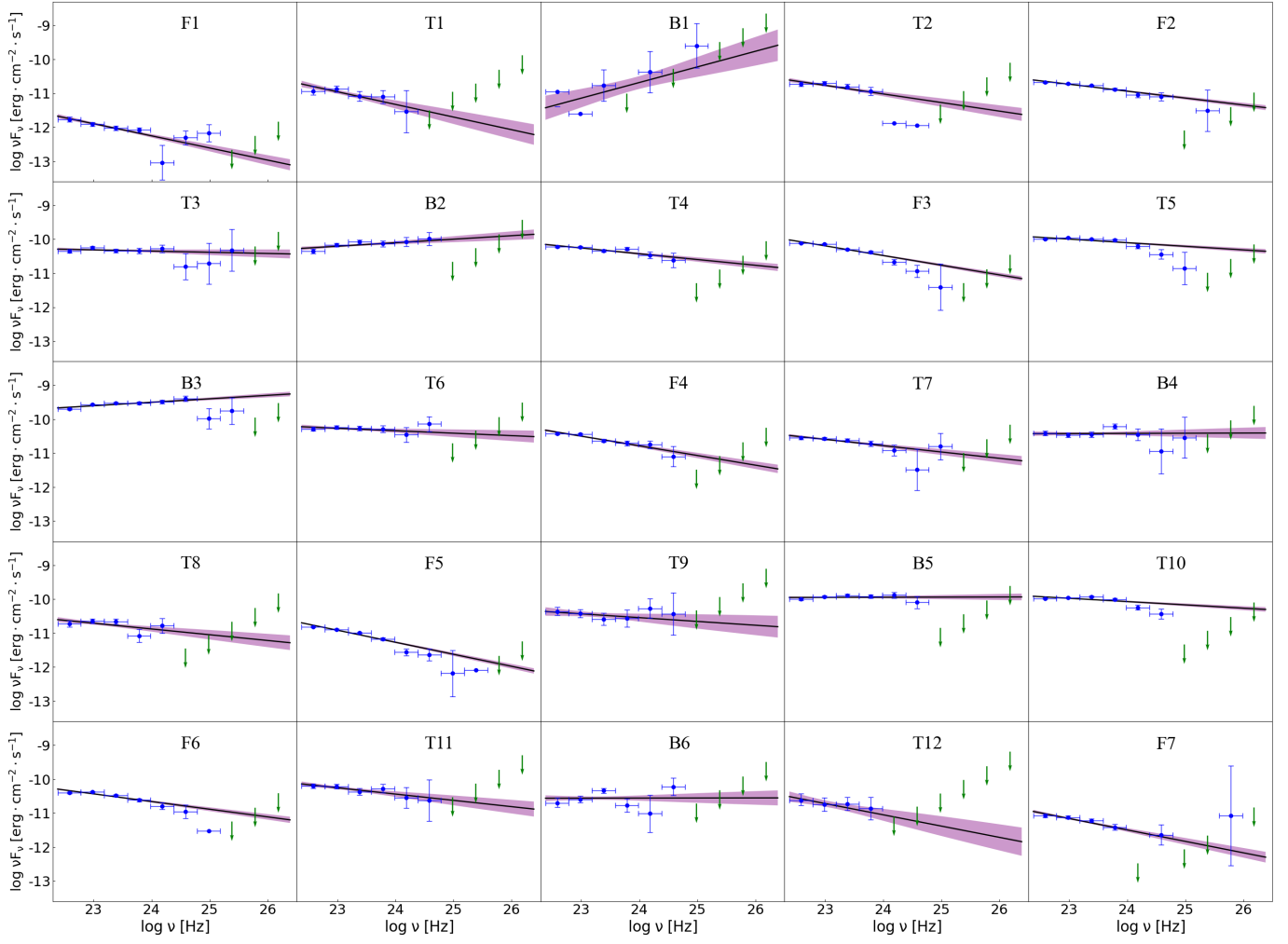}%fermi-sed
		\centering
	\caption{\textit{Fermi}-LAT SED for the MJD 54628-60387 epoch. The solid black line represents the best fit from the likelihood analysis over the entire energy range. Blue dots denote $\gamma$-ray data points, while green downward arrows indicate the upper energy limits. The purple area represents the 1$\sigma$ confidence band. The `letters + numbers' in each subgraph correspond to the states listed in Table~\ref{tab:1}. The corresponding images for the epochs F1, T1, B1, T2, F2, T3, and B2 are from \cite{2024A&A...685A.140R}.} 
	\label{Fig_index_FTB}          
\end{figure*}
%-------------------------------------------------------------------

We processed the light curves of MJD 54628-60387 (shown in Figure~\ref{Fig01}).
Figure~\ref{Fig01} depicts the location of the CL phenomenon (MJD 54628-58677; \citealt{2024A&A...685A.140R}), with shades of purple, green, and gray representing the FSRQ state (F epoch), BL Lac state (B epoch), and transition state (T epoch), respectively.
Post-CL phenomenon, the light variation remains notably dynamic. A significant correlation is observed between the photon spectral index of $\gamma$-ray $\Gamma_{\gamma}$ (calculated using \textit{Fermipy}), and the EW of Mg \uppercase\expandafter{\romannumeral2} during CL phenomena (as measured by \cite{2021ApJ...913..146M}, with $\Gamma_{\gamma}$ corresponding closely to the same epochs as the EW (refer to Figure~\ref{Fig_index_EW}).
Figure~\ref{Fig_index_EW} illustrates that $\Gamma_{\gamma}$ is relatively small during the BL Lac state, correlating with a relatively small EW.
Conversely, during the FSRQ state $\Gamma_{\gamma}$ is relatively large corresponding to a larger EW.
Comparison between FSRQ and BL Lac state reveals a division at
$\Gamma_{\gamma}$ $\sim$ 2.0, consistent with findings by \cite{2010ApJ...710.1271A}, which distinguish BL Lacs and FSRQs based on the $\gamma$-ray spectral index $\Gamma_{\gamma}$  $\sim$  2.2. 
Building on these results and theories, we hypothesize that OQ 344 continues in the CL state post MJD 54628-58677.
Hence, we define the F epoch as $\Gamma_{\gamma} $ $\gtrsim$ 2.2 indicating the FSRQ state; the B epoch as 
$\Gamma_{\gamma}$ $\lesssim$ 2.0  indicating the BL Lac state; and the T epoch as $2.0 < \Gamma_{\gamma}  <  2.2$ indicating the transition state.

We construct time intervals and $\gamma$-ray spectra to investigate the source's state evolution across MJD 58678-60387.
Comparisons of these spectra reveal significant differences across epochs (refer to Figure~\ref{Fig_index_FTB}).
The spectral index $\Gamma_{\gamma}$, summarized in Table ~\ref{tab:1}, categorizes into three distinct classes: $\Gamma_{\gamma} $ $\gtrsim$ 2.2 corresponds to the FSRQ state (i.e., ${F_3}$, ${F_4}$, ${F_5}$, ${F_6}$, and ${F_7}$), $\Gamma_{\gamma}$ $\lesssim$ 2.0 denotes the BL Lac state (i.e., ${B_3}$, ${B_4}$, ${B_5}$ and ${B_6}$), and $2.0 < \Gamma_{\gamma}  <  2.2$ indicates the transition state (i.e., ${T_4}$, ${T_5}$, ${T_6}$, ${T_7}$, ${T_8}$, ${T_9}$, ${T_{10}}$, ${T_{11}}$, and ${T_{12}}$).

%----------------------

\subsection{ Multiwavelength SED }
Here, we explore the variability across different epochs using the accretion rate as a test parameter.
In this work, we employ the conventional one-zone Syn+ SSC+ EC model to fit the SEDs of epochs${T_4}$, ${F_3}$, ${T_5}$, ${B_3}$, ${T_6}$, ${F_4}$, ${T_7}$, ${B_4}$, ${T_8}$, ${F_5}$, ${T_9}$, ${B_5}$, ${T_{10}}$, ${F_6}$, ${T_{11}}$, ${B_6}$, ${T_{12}}$, and ${F_7}$ (refer to Figure~\ref{Fig_sed}).
To refine the SED fitting for these 18 epochs, we gathered simultaneous multiwavelength data (refer to Table~\ref{tab:1}).
Due to the flux errors in the $g$-band data of ASAS-SN being too discrete,
%Due to the discrete flux errors in the g-band data from ASAS-SN,
only the flux values are retained, and the arithmetic average of $g$-band flux data per epoch serves as the simultaneous data point.
During the fitting process, red and blue dots denote simultaneous and nonsimultaneous data points, respectively. Where $g$-band data are insufficient, such as in the ${F_7}$  epoch, archival data are used to constrain the Syn component.

For epochs lacking simultaneous optical/UV or X-ray data, average data from all Swift observations in MJD 58678-60387 are used as nonsimultaneous data points.
The results also include the accretion rate  and the jet radiation power across different epochs (see Figure~\ref{Fig_jet}). 
For detailed information on the selection and fitting of observed data points in the SED, as well as the 1$\sigma$ parameter space, please refer to \cite{2024A&A...685A.140R} for comprehensive details and references.

%--------------------------------------------------------
%\begin{figure}[h]
    %\centering
    %\begin{minipage}[t]{0.48\textwidth}
        %\centering
        %\includegraphics[width=\linewidth]{Figure/T4.pdf}
        %\caption{BL Lac conversion to FSRQ state in MJD 58678-58728}
        %\label{fig:a}
    %\end{minipage}
    %\hfill % 可选的间距
    %\begin{minipage}[t]{0.48\textwidth}
        %\centering
        %\includegraphics[width=\linewidth]{Figure/F3.pdf}
        %\caption{The FSRQ state in MJD 58729-58806}
        %\label{fig:b}
    %\end{minipage}
%\end{figure}
%----------------------------------

%-------------------------------------------------
\begin{figure*}[htbp]
	\centering
	\subfigbottomskip=2pt
	\subfigcapskip=-5pt
	%\begin{adjustwidth}{-0.0cm}{1cm}
		\subfigure[\label{fig:a}][BL Lac conversion to FSRQ state in MJD 58678-58728]{
			\includegraphics[width=0.46\linewidth]{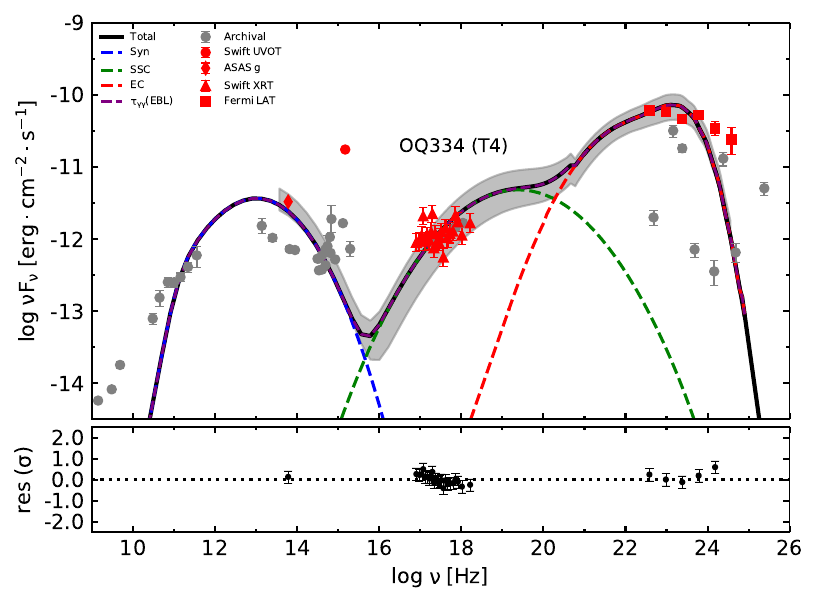}}\hspace{-1mm}
		\subfigure[\label{fig:b}][The FSRQ state in MJD 58729-58806]{
			\includegraphics[width=0.46\linewidth]{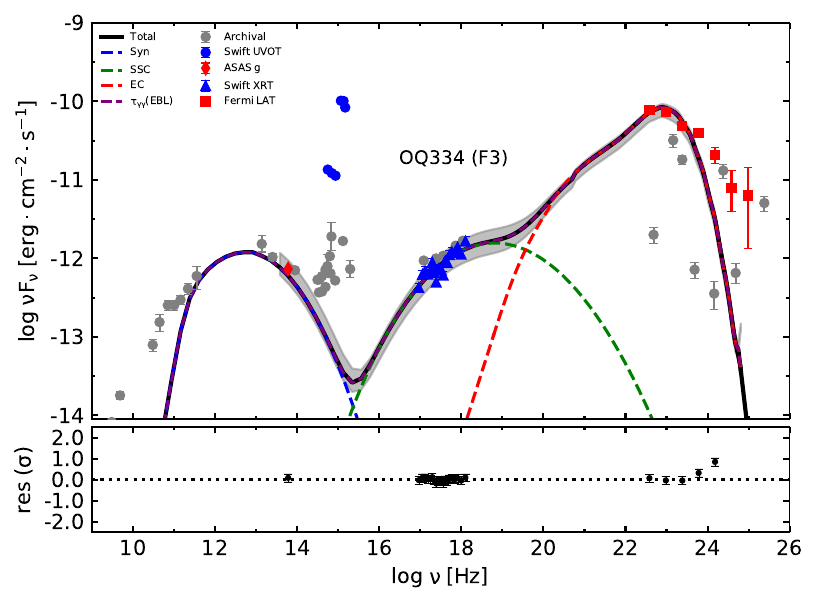}}\hspace{-1mm}
		\subfigure[\label{fig:c}][FSRQ conversion to BL Lac state in MJD 58807-58864]{
			\includegraphics[width=0.45\linewidth]{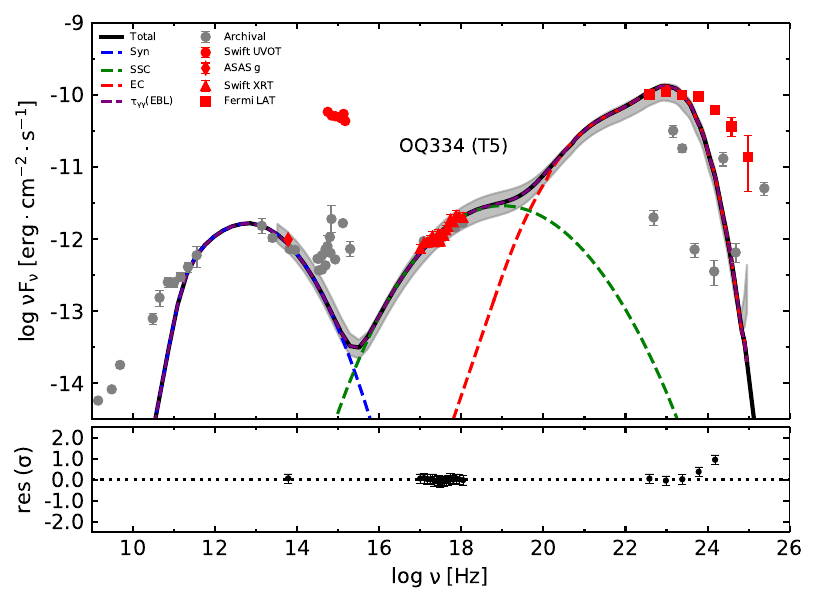}}
		\subfigure[\label{fig:d}][The BL Lac state in MJD 58865-58881]{
			\includegraphics[width=0.45\linewidth]{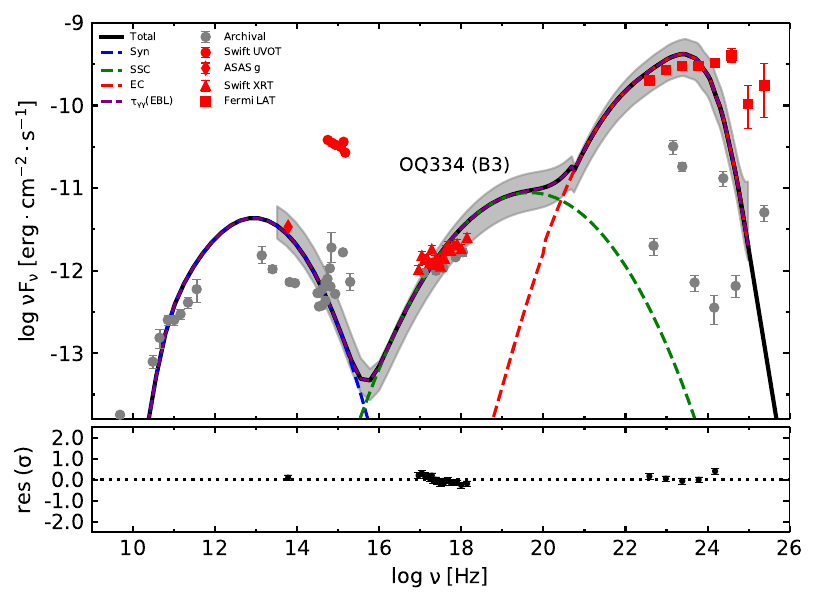}}\hspace{-1mm}
		\subfigure[\label{fig:e}][BL Lac conversion to FSRQ state in MJD 58882-58896]{
			\includegraphics[width=0.45\linewidth]{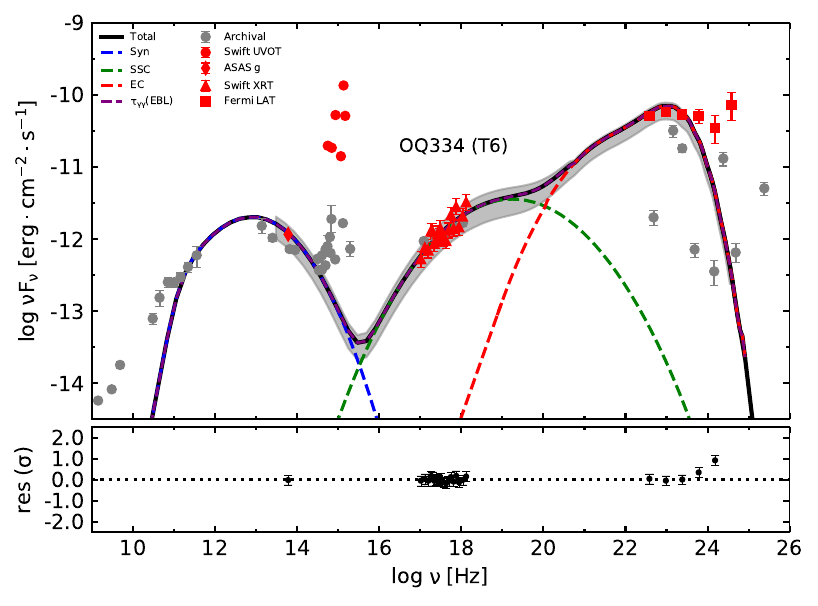}}\hspace{-1mm}
		\subfigure[\label{fig:f}][The FSRQ state in MJD 58897-58980]{
			\includegraphics[width=0.45\linewidth]{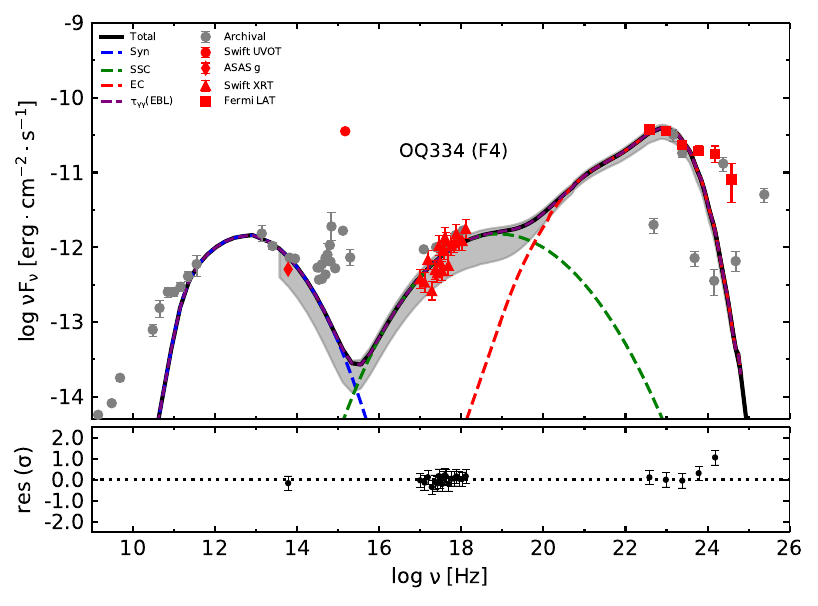}}
\end{figure*}

\begin{figure*}[htbp]
        \centering
	\subfigbottomskip=2pt
	\subfigcapskip=-5pt
		\subfigure[\label{fig:g}][FSRQ conversion to BL Lac state in MJD 58981-59042]{
			\includegraphics[width=0.45\linewidth]{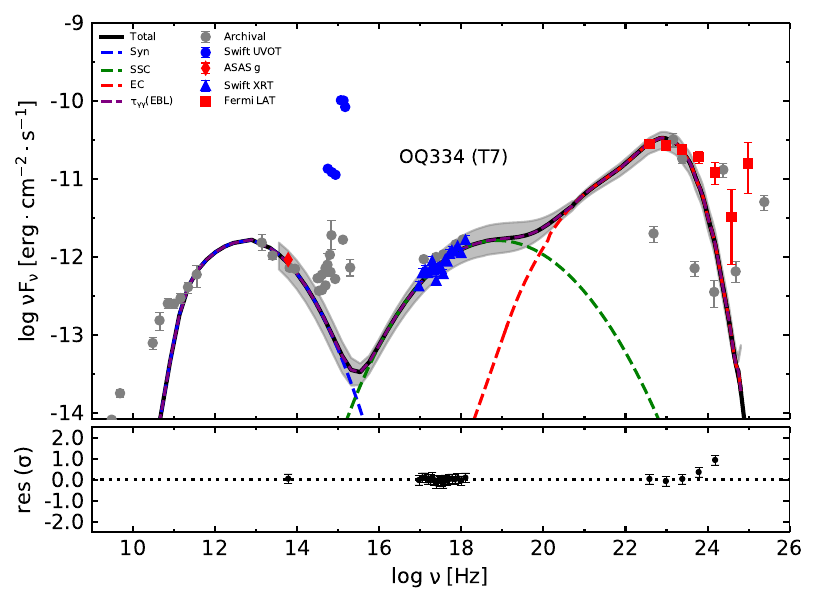}}\hspace{-2mm}
		\subfigure[\label{fig:h}][The BL Lac state in MJD 59043-59056]{
			\includegraphics[width=0.45\linewidth]{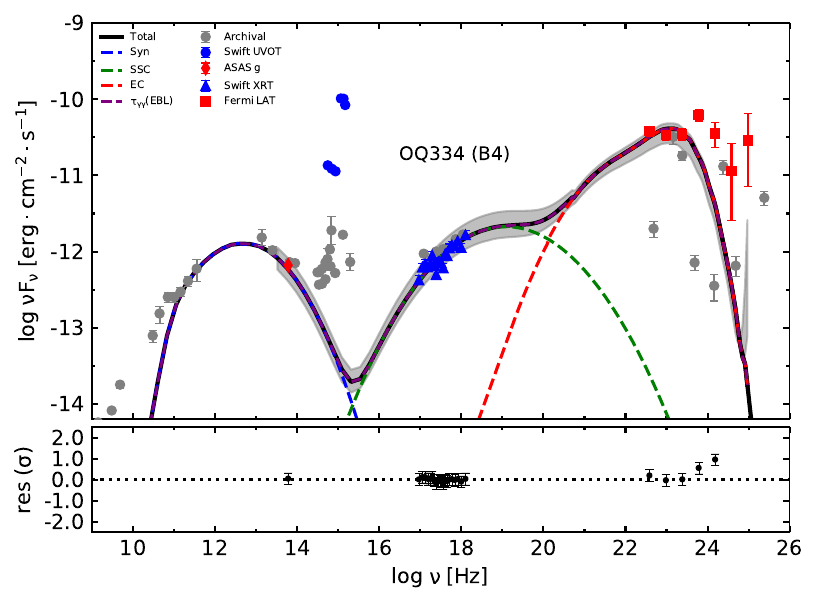}}\hspace{-1mm}
		\subfigure[\label{fig:i}][BL Lac conversion to FSRQ state in MJD 59057-59089]{
			\includegraphics[width=0.45\linewidth]{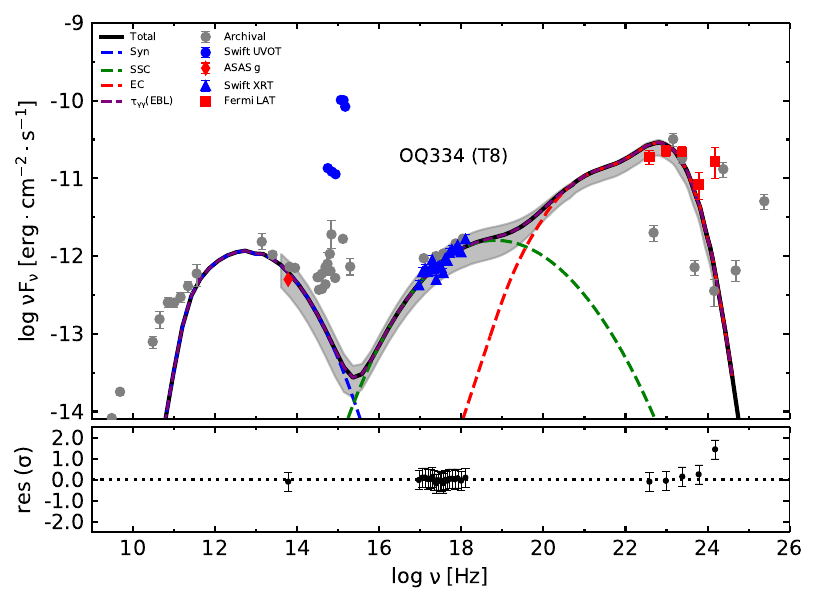}}
            \subfigure[\label{fig:j}][The FSRQ state in MJD 59090-59857]{
			\includegraphics[width=0.45\linewidth]{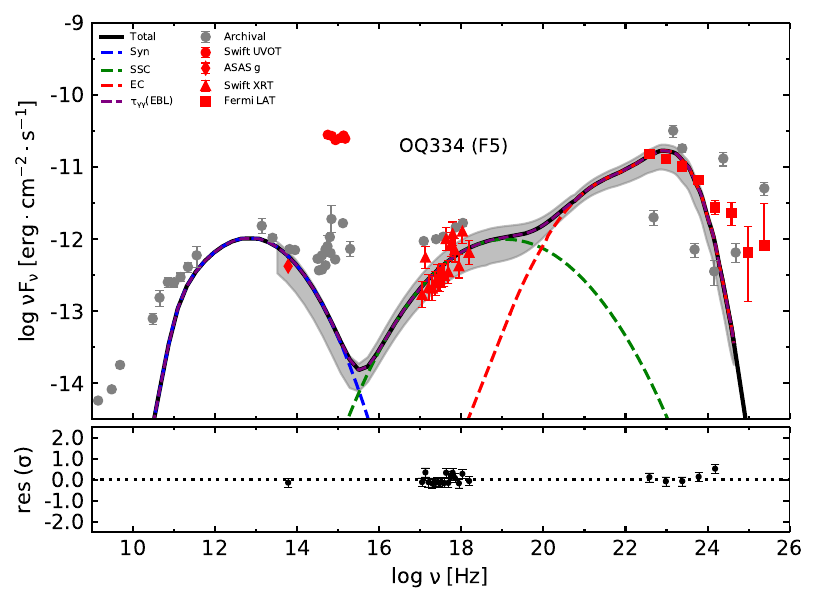}}\hspace{-1mm}
		\subfigure[\label{fig:k}][FSRQ conversion to BL Lac state in MJD 59858-59862]{
			\includegraphics[width=0.45\linewidth]{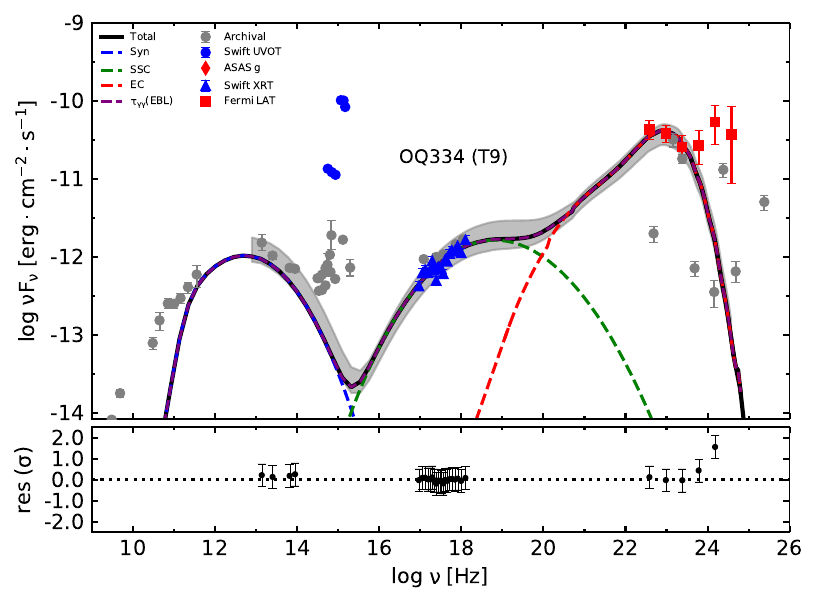}}\hspace{-1mm}
		\subfigure[\label{fig:l}][The BL Lac state in MJD 59863-59880]{
			\includegraphics[width=0.45\linewidth]{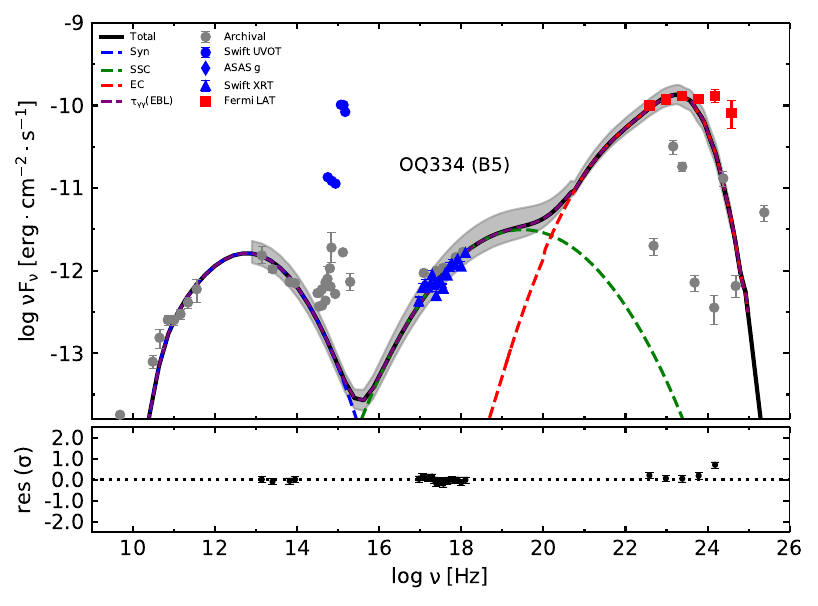}}\hspace{-1mm}
        %\end{adjustwidth}
\end{figure*}

\begin{figure*}[htbp]
	\centering
	\subfigbottomskip=2pt
	\subfigcapskip=-5pt
        %\begin{adjustwidth}{-0.0cm}{1cm}
		\subfigure[\label{fig:m}][BL Lac conversion to FSRQ state in MJD 59881-59931]{
			\includegraphics[width=0.45\linewidth]{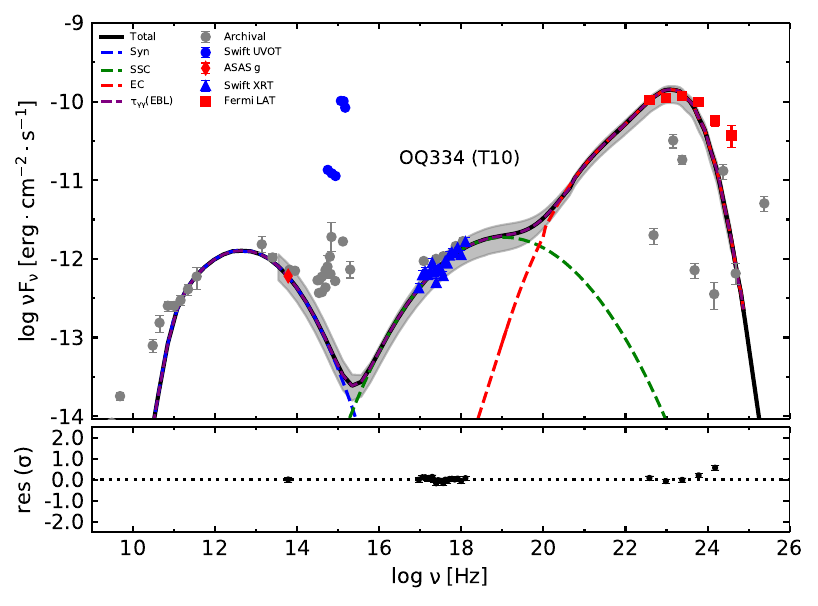}}\hspace{-1mm}
		\subfigure[\label{fig:n}][The FSRQ state in MJD 59932-60051]{
			\includegraphics[width=0.45\linewidth]{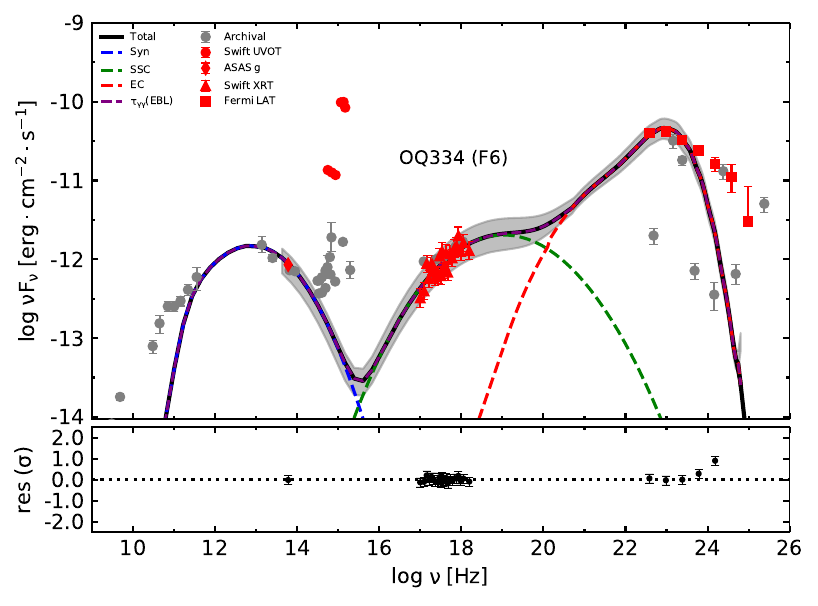}}\hspace{-1mm}
		\subfigure[\label{fig:o}][FSRQ conversion to BL Lac state in MJD 60052-60062]{
			\includegraphics[width=0.45\linewidth]{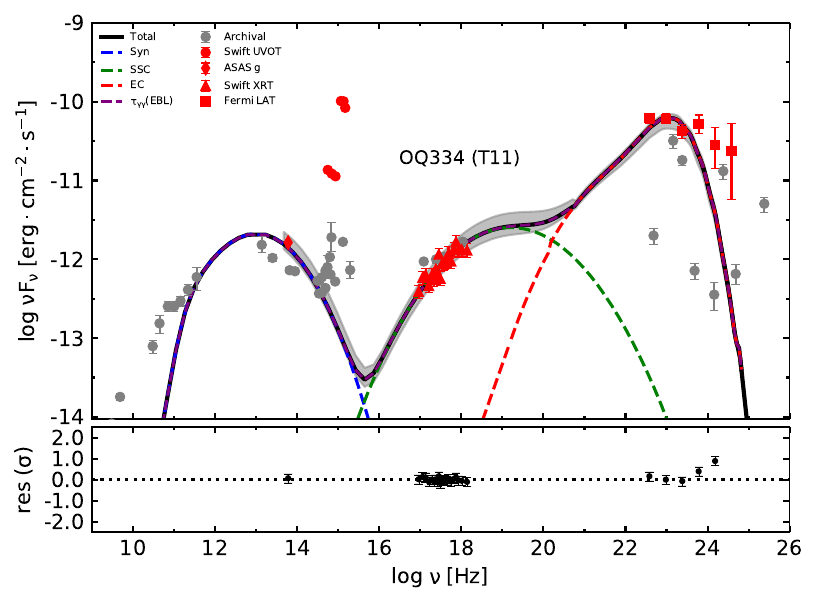}}
		\subfigure[\label{fig:p}][The BL Lac state in MJD 60063-60074]{
			\includegraphics[width=0.45\linewidth]{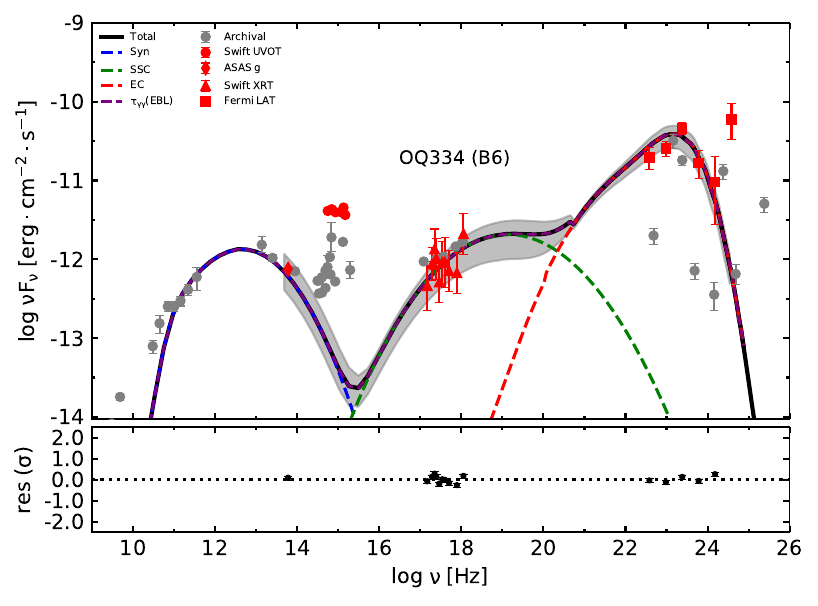}}\hspace{-1mm}
		\subfigure[\label{fig:q}][BL Lac conversion to FSRQ state in MJD 60075-60081]{
			\includegraphics[width=0.45\linewidth]{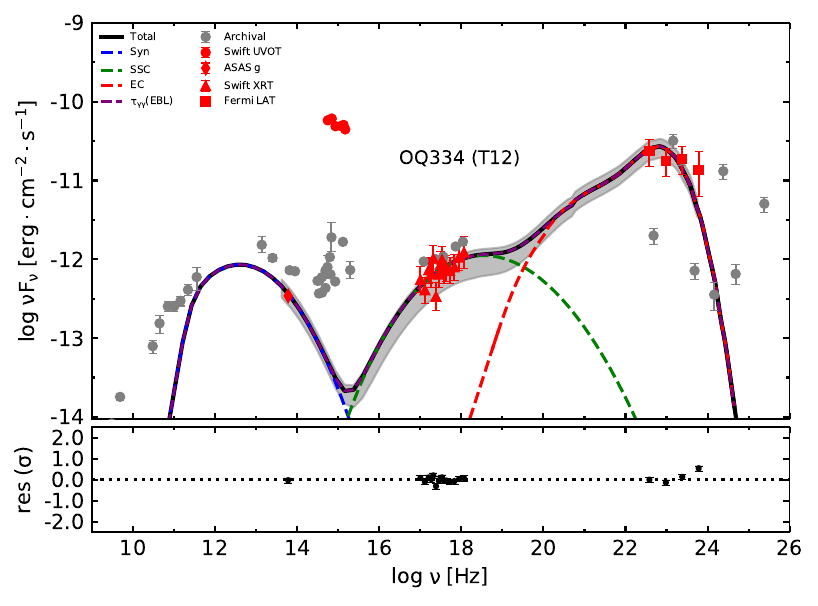}}\hspace{-1mm}
		\subfigure[\label{fig:r}][The FSRQ state in MJD 60082-60387]{
			\includegraphics[width=0.45\linewidth]{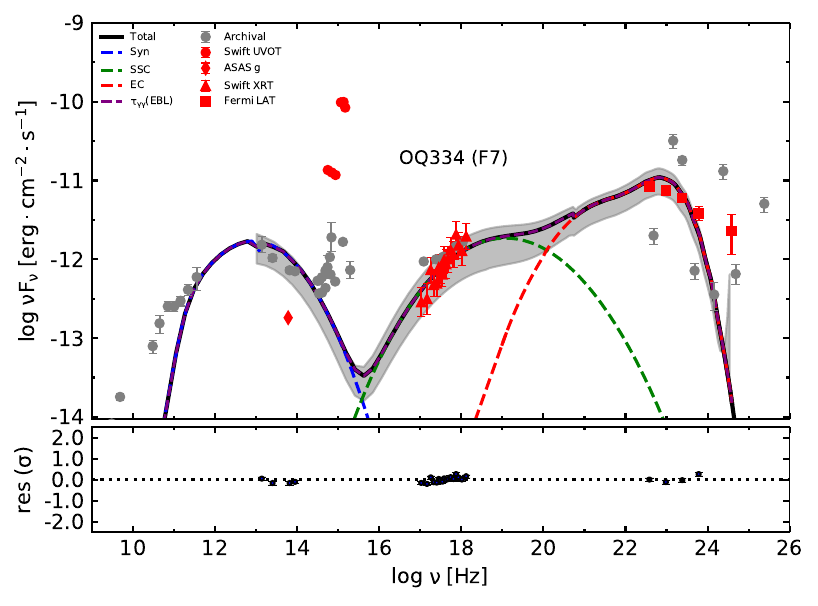}}
	%\end{adjustwidth}
	\caption{Multi-wavelength SED of the CLB across different states during MJD 54628-60387. The dotted blue, green, red, and purple lines represent Syn, SSC, EC, and EBL absorption components, respectively. The solid black line represents the total fitted spectrum of the SED.Red dots denote quasi-simultaneous data, blue dots represent non-quasi-simultaneous data, and grey dots indicate archival data. Each subplot corresponds to epochs T4, F3, T5, B3, T6, F4, T7, B4, T8, F5, T9, B5, T10, F6, T11, B6, T12, and F7 during MJD 54628-60387. The grey background shows the 1$\sigma$ parameter space of the SED model, starting from the optical observation data points used in the fit. The lower panel of each subplot displays the residuals of the fit. The SED of different states before the T4 epoch (i.e., MJD54628-58677) have been fitted in \cite{2024A&A...685A.140R}.}
	\label{Fig_sed}
\end{figure*}
%------------------------------------------------

%-------------------------------------------------------------------
\begin{figure}
	\centering
		\includegraphics[width=0.8\linewidth]{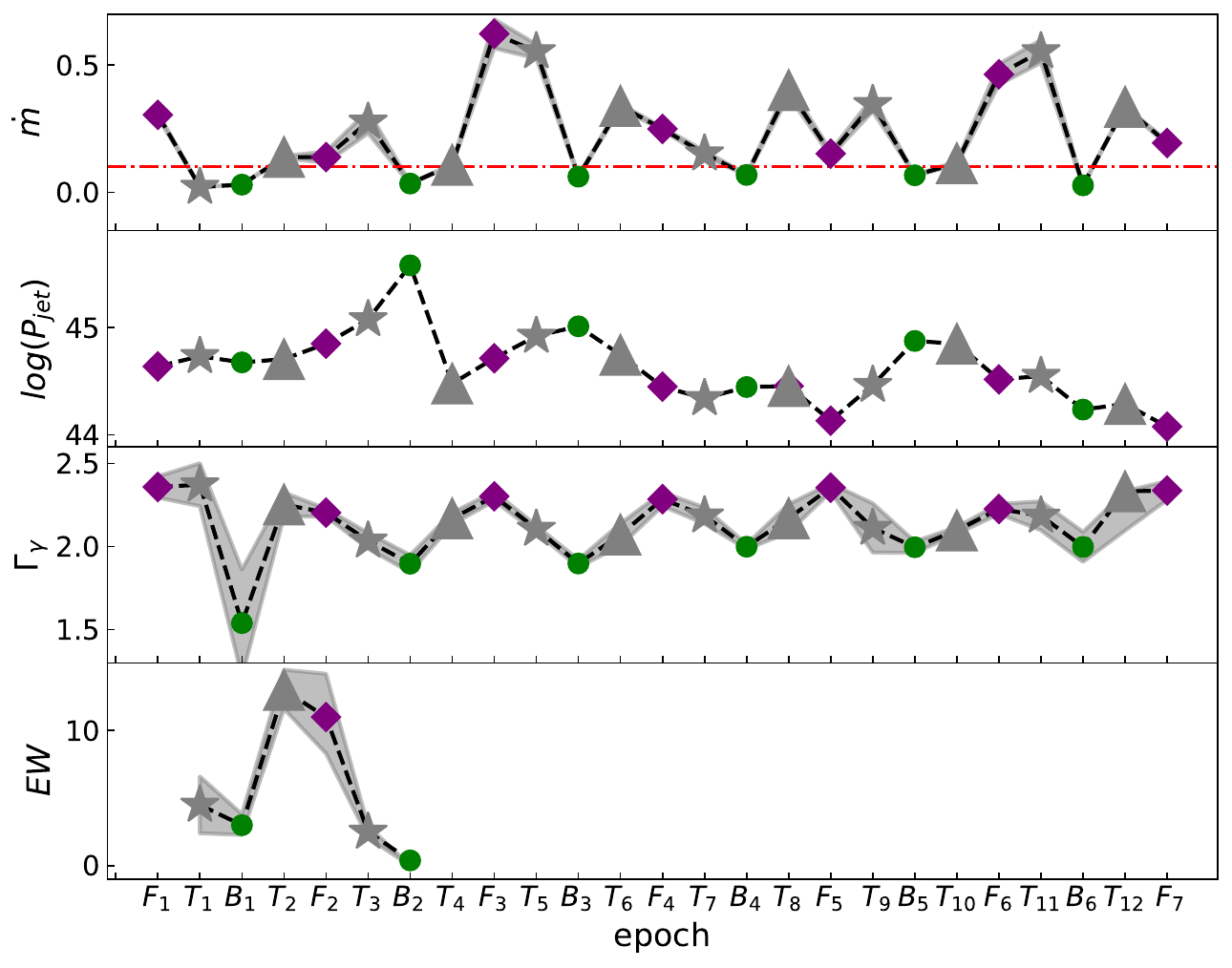}%accretion rate
		\centering
	\caption{ The accretion rates \.{m}, jet radiation power $log(P_{jet})$, photon spectral indices $\Gamma_\gamma$, and equivalent widths (EW) for epochs F, B, and T between MJD 54628 and 60387. In the top panel, the red dotted line indicates \.{m} = 0.1. Purple diamonds and green dots represent the FSRQ (F) and BL Lac (B) epochs. Gray pentagrams and triangles represent the transition eopchs (T) from the FSRQ state to the BL Lac state and from the BL Lac state to the FSRQ state, respectively. The gray area represents the 1$\sigma$ confidence band. It is worth noting that due to the large errors during the T12 epoch, the errors of the T12 epoch are neglected when considering the 1$\sigma$ confidence band. \.{m} and $log(P_{jet})$ of these epochs that are epochs F1, T1, B1, T2, F2, T3, and B2 have been re-fitted using $H_{0}=73.3$ km s$^{-1}$ Mpc$^{-1}$. } 
	\label{Fig_jet}          
\end{figure}
%------------------

%-------------------------------------------------------------------
\begin{figure}
	\centering
		\includegraphics[width=0.8\linewidth]{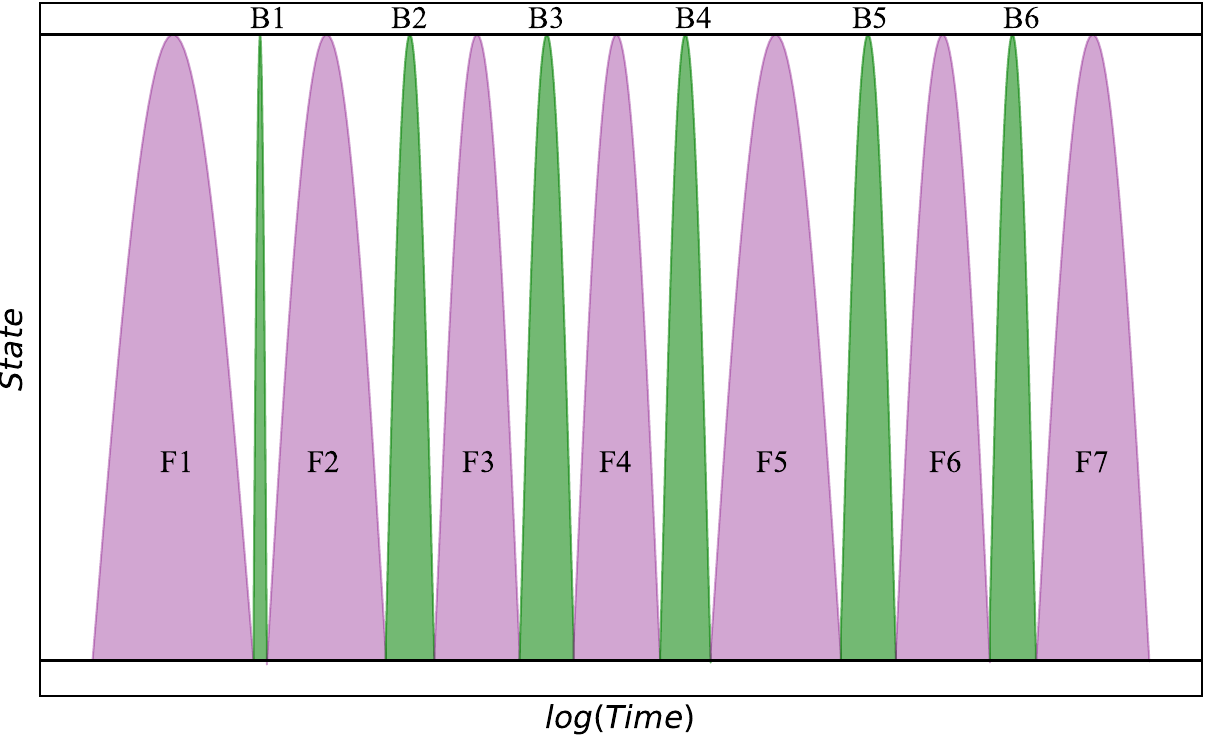}%state_2
		\centering
	\caption{Schematic of the changing-look blazar (CLB) long-term evolution. Green shading indicates the BL Lac state, while purple represents the FSRQ state. The vertical axis represents states, and the horizontal axis denotes the logarithmic values of the timescales for the epochs F and B. Data for the ${F_1}$, ${B_1}$, ${F_2}$, and ${B_2}$ are from \cite{2024A&A...685A.140R}.} 
	\label{Fig_time}          
\end{figure}

%-------------------------------------------------------------------

%-------------------------------------------------------------------
\begin{figure}
	\centering
		\includegraphics[width=0.8\linewidth]{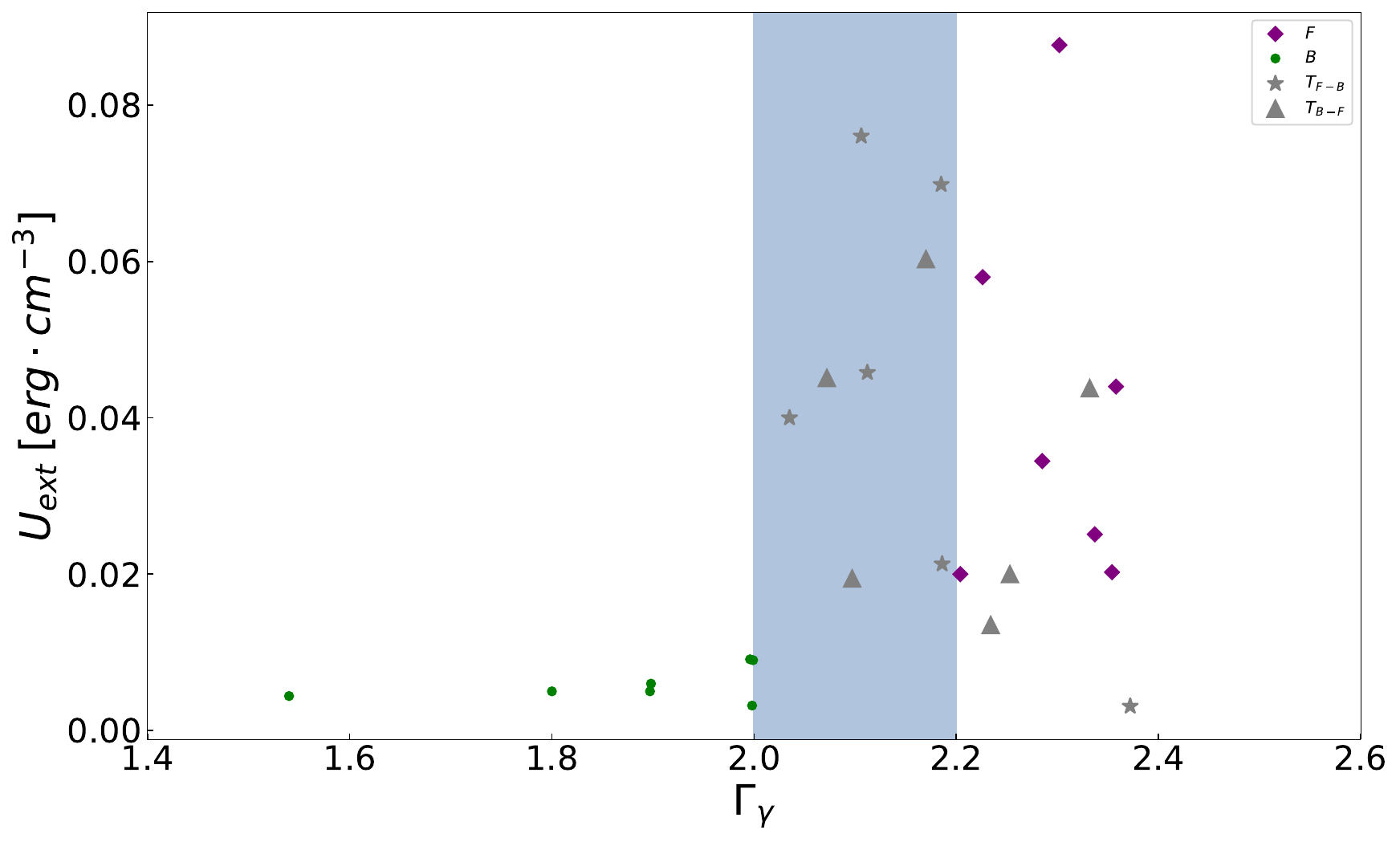}%U_ext——gamma
		\centering
	\caption{ External photon field energy density versus photon spectral index for epochs F, B, and T during MJD 54628-60387. Green dots and purple prisms correspond to values during the FSRQ (F) and BL Lac (B) epochs, respectively. Gray pentagrams and triangles represent values during the transition epoch (T) from FSRQ to BL Lac and BL Lac to FSRQ, respectively. The Light steel blue shading indicates epochs within the range of $\Gamma_{\gamma} = 2.0$ and $\Gamma_{\gamma} = 2.2$. The external photon field energy density for different states before the T4 epoch (i.e., MJD54628-58677) has been refitted according to $H_{0}=73.3$ km s$^{-1}$ Mpc$^{-1}$.} 
	\label{Fig_uext}          
\end{figure}

\section{Discussion}\label{sec:diss}
%-------------------------------------------------------------------
\subsection{\textit{Fermi} SED and Evolution of OQ 334}
%-------------------------------------------------------------------
Post MJD 54628-58677, the $\gamma$-ray flux remains active, suggesting that the source may still be in a changing-look phase.
Figure~\ref{Fig_index_EW} illustrates the correlation between the $\gamma$-ray photon spectral index and the EW of Mg \uppercase\expandafter{\romannumeral2}, indicating that spectral softness or hardness corresponds to varying emission line characteristics. 
Furthermore, Figure~\ref{Fig_index_FTB} highlights differences in spectral indices across different epochs, likely due to varying cooling rates of relativistic electrons and distinct EC components. These EC components influence the broad emission lines' strength and enable CLBs to display the characteristics of both FSRQs and BL Lacs during their evolutionary process.

We summarize the timescales of FSRQ and BL Lac states, plotted in Figure~\ref{Fig_time}, revealing irregular durations in the FSRQ state and an increasing trend in the BL Lac state. This trend suggests a potential evolution where an FSRQ may transition into a BL Lac object.

\subsection{Multiwavelength SED and CLB evolution}
\subsubsection{Accretion Rate and CLB Evolution}
Crucial roles in blazar classification, where besides the EW, $L_{disk}/L_{Edd}<10^ {-2}$ (\citealt{2010MNRAS.402..497G}) is a defining criterion.
In this work, we denoted $L_{disk}/L_{Edd}$ as \.{m} in Table ~\ref{tab:2}.
Previous research (\citealt{2024A&A...685A.140R}) indicates that accretion rates are $>$ 0.1 in both FSRQ and transition states and $<$ 0.1 in the BL Lac state.
Tables \ref{tab:1} and \ref{tab:2} reveal that $\Gamma_{\gamma}$ $\lesssim$ 2.0 corresponds to $L_{disk}/L_{Edd}<10^ {-1}$, indicative of the BL Lac state. 
Conversely, $\Gamma_{\gamma} $ $\gtrsim$ 2.2 corresponds to $L_{disk}/L_{Edd}>10^ {-1}$, suggesting the FSRQ state.
These findings underscore the relationship between $\gamma$-ray spectrum characteristics and accretion rate magnitude.
The classification into F and B epochs mirrors the approach of \cite{2009MNRAS.396L.105G} and \cite{2011MNRAS.414.2674G}, where the $\gamma$-ray spectral index distinguishes between FSRQs and BL Lacs. 
PMN J2345-1555 (\citealt{2013MNRAS.432L..66G}) also exhibits behavior similar to OQ 334, where the SED varies across different epochs. \cite{2013MNRAS.432L..66G} suggest that during the process of the radiation region moving from within the BLR to outside it, the radiation energy density decreased, and cooling reduced. Electrons could reach higher energies, leading to the spectral energy distribution of PMN J2345-1555 changing from ``red'' to ``blue''. In our work, during the state transition from FSRQ to BL Lac, the location of the radiation region moved outward, and the energy density of the external photon field decreased in this process, allowing relativistic electrons to be accelerated to higher energies, which is similar to the results of \cite{2013MNRAS.432L..66G}. However, the difference is that the location of the radiation region is smaller than the outer radius of the BLR, which is $8.19\times 10^{17}cm$ (for details, see \citealt{2024A&A...685A.140R}). The radiation region moves within the BLR and does not move outside it.
Generally, BL Lacs exhibit harder $\gamma$-ray spectra and lower luminosity than FSRQs, potentially influenced by diverse accretion models (\citealt{2009MNRAS.396L.105G}).
When the CLB's accretion rate drops below 0.1 (\citealt{1995ApJ...444..231N,1996ApJ...462..142L,1999ApJ...516..177G}), the advection-dominated accretion flow (ADAF) dominates, limiting ionization within the BLR clouds (For detailed descriptions please refer to \citealt{2024A&A...685A.140R}) 
This results in weakened broad emission lines, characteristic of the BL Lac state.
Lower accretion rates reduce $\gamma$-ray production via the weaker EC component, resulting in a smaller $\gamma$-ray photon spectral index observed in the BL Lac state compared to the FSRQ state.

\subsection{The CLB and Blazars Evolution}

The jet power during epochs B2, B3, and B5 (Figure~\ref{Fig_jet}) is stronger than that in the FSRQ state, which is different from the general situation where the overall jet power of the vast majority of FSRQs is greater than that of BL Lacs (\citealt{2018ApJS..235...39C}). This may be because we consider using the model Syn + SSC + EC (\citealt{2024A&A...685A.140R}) to fit the SED in the BL Lac state. In the FSRQ state, the accretion rate has a similar trend when the EW increases. However, due to the lack of EW data, the change in EW may also be caused by the combined effect of jet power and accretion rate.

Accretion in FSRQs typically involves a thin and hot standard accretion disk (SSD), whereas BL Lac objects often exhibit accretion through an ADAF. 
We observe distinct accretion rates between FSRQ and BL Lac states, likely due to these different accretion models (\citealt{2024A&A...685A.140R}).
Thus, from the perspective that changes in accretion rates drive blazar evolution, the transition from an FSRQ to a BL Lac object essentially involves a shift from an SSD to an ADAF.
\cite{2002ApJ...571..226C} also noted this transition. They suggested a decrease in accretion rate during the evolution from FSRQs to BL Lac objects. This evolutionary process of the accretion disk undergoes a lengthy and intricate transformation reminiscent of the accretion model transition observed in CLBs from the FSRQ to the BL Lac state.
The similar transitions observed suggest that CLBs may represent a special transitional phase in blazar evolution.
We also observe variations in the intensity of the external photon field, indicating that CLBs may occupy a distinct stage in blazar evolution. 
From the external photon field perspective, we expect a decrease in energy density during the evolution from FSRQs to BL Lac objects. 
Although the external photon field energy density fluctuates over extended epochs (see Figure~\ref{Fig_uext} ), the data reveal a significantly higher energy density in the FSRQ state compared to the BL Lac state. 
Moreover, the FSRQ state correlates with a higher photon spectral index and greater external photon field energy density.
Meanwhile, the BL Lac state exhibits a lower spectral index and reduced external photon field energy density.
These differences likely stem from the varying accretion modes associated with each state.

During CLB evolution, substantial divergence in both external photon field energy density and photon spectral index occurs due to changes in accretion modes. This phenomenon mirrors broader blazar evolutionary trends. Therefore, we propose that blazar evolution is not a straightforward progression from FSRQs to BL Lacs but a complex and protracted process. Transitional stages as CLBs may represent critical phases within this evolutionary trajectory.

%---------------------------------------

%-------------------------------------------------------------------
\section{Conclusion}   

The $\gamma$-ray photon spectral index in the CLB can be categorized into three distinct states: $\Gamma_{\gamma} $ $\gtrsim$ 2.2 (FSRQ state: ${F_3}$, ${F_4}$, ${F_5}$, ${F_6}$, and ${F_7}$), $\Gamma_{\gamma}$ $\lesssim$ 2.0 (BL Lac state: ${B_3}$, ${B_4}$, ${B_5}$, and ${B_6}$), and $2.0 < \Gamma_{\gamma}  <  2.2$ (Transition state: ${T_4}$, ${T_5}$, ${T_6}$, ${T_7}$, ${T_8}$, ${T_9}$, ${T_{10}}$, ${T_{11}}$, and ${T_{12}}$)  as detailed in Table~\ref{tab:1} and Figure~\ref{Fig_index_FTB}. 
We analyzed Fermi-LAT and multiwavelength SEDs across these 18 epochs. Our findings are summarized as follows: (1) There is a clear correlation between $\Gamma_{\gamma}$ and the EW in the FSRQ and BL Lac states. (2) The timescale for the FSRQ state does not exhibit a consistent decreasing trend, whereas the BL Lac state shows a clear increasing trend. (3) OQ 334 may still be undergoing a changing-look process after the previously observed phenomenon in MJD 54628-58677. (4) The CLB may represent a unique stage in the evolutionary process of blazars.

In summary, these results strongly suggest that the CLB variant is a distinct and significant phase closely tied to the evolution of blazars.

%----------------------------------------------

\begin{sidewaystable*}[!htp]
	\tablenum{1}
	\centering
	\small
	\begin{center}
		\caption{\textit{Fermi}-LAT, \textit{Swift}, ASAS-SN, and {Fermi}-LAT SED results in MJD 54628-60387}
            \label{tab:1}
		\setcounter{table}{1}
		\renewcommand{\thetable}{5/arabic{table}}
		\renewcommand\arraystretch{1.02}
		\begin{adjustwidth}{-2cm}{0cm}
			\scalebox{1.01}{
                 \begin{tabular}{cccccccc} % four columns, alignment for each
		\hline \hline
        Epoch         & State         &\textit{Fermi}-LAT & \textit{Fermi}-LAT   & $\Gamma_{\gamma}$ & Swift      & ASAS-SN \\
         ...                                        &  ...                                        & MJD                                       & year/month/day  &...       & Obsid (X-ray\& Optical)        & UT Date ($g$-band)\\
         (1)                                        &  (2)                                        & (3)                                       & (4)            &(5)       & (6)        & (7)  \\
		\hline
					\centering
					 ${T_4}$	&	BL Lac$\longrightarrow$FSRQ	&	58678-58728	&	2019/7/14-2019/9/2	&		2.17$\pm$0.04		&	11,12	&	2019/07/14.2647123-2019/09/02.0922228\\	
${F_3}$	&	FSRQ	&	58729-58806	&	2019/9/3-2019/11/19	&		2.30$\pm$0.03		&	$\ast$	&	2019/09/03.0962627-2019/09/05.0970780	\\
${T_5}$	&	FSRQ$\longrightarrow$BL Lac	&	58807-58864	&	2019/11/20-2020/1/16	&		2.11$\pm$0.02		&	13,14,15,16,17	&	2019/12/08.5322031-2020/01/12.5386304	\\
${B_3}$	&	BL Lac	&	58865-58881	&	2020/1/17-2020/2/2	&		1.90$\pm$0.02		&	18,19,22,23,24,25,26,27,28	&	2020/01/19.4329803-2020/02/02.4206879\\	
${T_6}$	&	BL Lac$\longrightarrow$FSRQ	&	58882-58896	&	2020/2/3-2020/2/17	&		2.07$\pm$0.07		&	29,30,31	&	2020/02/03.4163189-2020/02/09.6177725\\	
${F_4}$	&	FSRQ	&	58897-58980	&	2020/2/18-2020/5/11	&		2.29$\pm$0.04		&	32,33,34	&	2020/02/24.4607216-2020/05/09.2806503\\	
${T_7}$	&	FSRQ$\longrightarrow$BL Lac	&	58981-59042	&	2020/5/12-2020/7/12	&		2.19$\pm$0.05		&	$\ast$	&	2020/05/13.2364417-2020/07/12.2838929\\	
${B_4}$	&	BL Lac	&	59043-59056	&	2020/7/13-2020/7/26	&		2.00$\pm$0.06		&	$\ast$	&	2020/07/14.3374582-2020/07/26.2197564\\	
${T_8}$	&	BL Lac$\longrightarrow$FSRQ	&	59057-59089	&	2020/7/27-2020/8/28	&		2.17$\pm$0.09		&	$\ast$	&	2020/07/28.1575469-2020/08/27.1341375	\\
${F_5}$	&	FSRQ	&	59090-59857	&	2020/8/29-2022/10/5	&		2.35$\pm$0.03		&	35,36,37	&	2020/08/29.1260993-2022/09/16.0851873	\\
${T_9}$	&	FSRQ$\longrightarrow$BL Lac	&	59858-59862	&	2022/10/6-2022/10/10	&		2.11$\pm$0.15		&	$\ast$	&	$\star$\\	
${B_5}$	&	BL Lac	&	59863-59880	&	2022/10/11-2022/10/28	&		2.00$\pm$0.03		&	$\ast$	&	$\star$\\	
${T_{10}}$	&	BL Lac$\longrightarrow$FSRQ	&	59881-59931	&	2022/10/29-2022/12/18	&		2.10$\pm$0.02		&	$\ast$	&	2022/11/24.5218769-2022/12/10.6299581\\	
${F_6}$	&	FSRQ	&	59932-60051	&	2022/12/19-2023/4/17	&		2.23$\pm$0.03		&	39,40,41,42,43	&	2022/12/19.4975361-2023/04/17.3163449\\	
${T_{11}}$	&	FSRQ$\longrightarrow$BL Lac	&	60052-60062	&	2023/4/18-2023/4/28	&		2.19$\pm$0.09		&	44,45,47,48,49,50,51,52,53,54,55	&	2023/04/18.4808241-2023/04/27.4043848\\	
${B_6}$	&	BL Lac	&	60063-60074	&	2023/4/29-2023/5/10	&		2.00$\pm$0.09		&	56	&	2023/04/29.2858593-2023/05/10.4073788\\	
${T_{12}}$	&	BL Lac$\longrightarrow$FSRQ	&	60075-60081	&	2023/5/11-2023/5/17	&		2.33$\pm$2.10		&	57,58,59	&	2023/05/11.3967560-2023/05/17.3534203\\	
${F_7}$	&	FSRQ	&	60082-60387	&	2023/5/18-2024/3/18	&		2.34$\pm$0.06		&	60,61,62,63	&	2023/05/22.3352880-2024/03/17.5884172\\	
					\hline
			\end{tabular}}\\
		\end{adjustwidth}
	\end{center} 
	\tablecomments{Note. Column (1) gives the nomenclature of the epochs in changing-look (CL); Column (2) indicates the states of CL blazar OQ 334 in various epochs; Columns (3)-(4) provide the epoch records of $\gamma$-ray observations in CL by \textit{Fermi}-LAT, expressed in units of MJD and year/month/day, respectively; Column (5) indicates the photon spectral index of $\gamma$-ray along with their 1$\sigma$ confidence interval; Columns (6) - (7) give the epoch records of the X-ray and UV/optical bands observed by Swift, respectively, where $\ast$ represents that data exists for this stage but not specifically from Swift, and $\star$ signifies that no corresponding data is found.  }		
	
\end{sidewaystable*}
%------------------------------------------

%-------------------------------
\begin{sidewaystable*}[!htp]
	\tablenum{2}
	\centering
	\small
	\begin{center}
		\caption{Multiwavelength SED fits local optimal parameters and the chi-square value $\chi ^{2}$.}
            \label{tab:2}
		\setcounter{table}{2}
		\renewcommand{\thetable}{5/arabic{table}}
		\renewcommand\arraystretch{1.02}
		\begin{adjustwidth}{-4cm}{0cm}
			\scalebox{1.01}{
                 \begin{tabular}{cccccccccccccc} % four columns, alignment for each
		\hline \hline
        Epoch &\makebox[0.01\textwidth][c]{s}  &\makebox[0.01\textwidth][c]{b}  &\makebox[0.01\textwidth][c]{$\gamma_{0}$}  &\makebox[0.01\textwidth][c]{$\delta$} &\makebox[0.01\textwidth][c]{B} &\makebox[0.01\textwidth][c]{R} & \makebox[0.01\textwidth][c]{$r_{dis}$} &\makebox[0.01\textwidth][c]{$U_{ext}$}& \makebox[0.01\textwidth][c]{$\chi^{2}$} & $L_{disk}$           & \.{m}  & $log(P_{jet})$\\
         ...   &...&...& $10^{3}$ &...&...& $10^{16}$ & $10^{16}$ & $10^{-2}$ & ... & $10^{46}$           &  ...\\
         ...   &...&...          & ...&...& G& cm & cm & $erg\cdot cm^{-3}$ &... & $erg\cdot s^{-1}$           &  ... &  $erg\cdot s^{-1}$\\
		\hline
       ${T_4}$&5.59$\pm$0.09 &1.54$\pm$0.01&7.52$\pm$0.11 &14.65$\pm$0.15 &0.26$\pm$0.01 &2.26$\pm$0.02 &2.33$\pm$0.15 &1.35$\pm$0.23 &1.11&0.47$\pm$0.08&0.09$\pm$0.02&44.48\\
${F_3}$&5.60$\pm$0.01 &1.38$\pm$0.01&7.11$\pm$0.17 &11.68$\pm$0.32 &0.41$\pm$0.03 &1.80$\pm$0.05 &1.67$\pm$0.10 &8.77$\pm$0.48 &0.14&3.04$\pm$0.17&0.61$\pm$0.03&44.71\\
${T_5}$&5.72$\pm$0.03 &1.47$\pm$0.01&7.82$\pm$0.24 &12.02$\pm$0.15 &0.31$\pm$0.01 &1.85$\pm$0.02 &1.82$\pm$0.04 &7.61$\pm$0.70 &0.06&2.64$\pm$0.24&0.53$\pm$0.05&44.92\\
${B_3}$&5.22$\pm$0.09 &1.59$\pm$0.01&8.83$\pm$0.14 &18.42$\pm$0.20 &0.08$\pm$0.01 &2.84$\pm$0.03 &4.62$\pm$0.03 &0.60$\pm$0.06 &0.5&0.21$\pm$0.02&0.04$\pm$0.01&45.01\\
${T_6}$&5.53$\pm$0.04 &1.47$\pm$0.01&7.98$\pm$0.13 &10.58$\pm$0.25 &0.31$\pm$0.02 &1.63$\pm$0.04 &2.55$\pm$0.04 &4.51$\pm$0.35 &0.23&1.56$\pm$0.12&0.31$\pm$0.02&44.74\\
${F_4}$&5.66$\pm$0.08 &1.47$\pm$0.03&6.96$\pm$0.23 &11.44$\pm$0.65 &0.32$\pm$0.01 &1.76$\pm$0.10 &1.99$\pm$0.05 &3.45$\pm$0.16 &0.46&1.19$\pm$0.56&0.24$\pm$0.01&44.45\\
${T_7}$&5.42$\pm$0.05 &1.35$\pm$0.01&6.79$\pm$0.29 &11.89$\pm$0.20 &0.33$\pm$0.01 &1.83$\pm$0.03 &2.29$\pm$0.08 &2.13$\pm$0.21 &0.14&0.74$\pm$0.07&0.15$\pm$0.01&44.34\\
${B_4}$&5.43$\pm$0.05 &1.47$\pm$0.01&7.68$\pm$0.12 &12.50$\pm$0.45 &0.14$\pm$0.01 &1.92$\pm$0.07 &3.22$\pm$0.18 &0.90$\pm$0.06 &0.16&0.31$\pm$0.02&0.06$\pm$0.01&44.45\\
${T_8}$&5.64$\pm$0.01 &1.37$\pm$0.01&7.31$\pm$0.15 &10.21$\pm$0.40 &0.55$\pm$0.04 &1.57$\pm$0.06 &1.51$\pm$0.05 &6.03$\pm$0.41 &0.16&2.09$\pm$0.14&0.42$\pm$0.03&44.45\\
${F_5}$&5.48$\pm$0.07 &1.40$\pm$0.03&7.10$\pm$0.15 &10.95$\pm$0.29 &0.32$\pm$0.03 &1.69$\pm$0.04 &2.00$\pm$0.19 &2.03$\pm$0.16 &0.72&0.70$\pm$0.06&0.14$\pm$0.01&44.13\\
${T_9}$&5.77$\pm$0.07 &1.43$\pm$0.01&7.59$\pm$0.13 &11.75$\pm$0.30 &0.33$\pm$0.04 &1.81$\pm$0.05 &1.45$\pm$0.05 &4.58$\pm$0.46 &0.15&1.59$\pm$0.16&0.32$\pm$0.03&44.46\\
${B_5}$&4.94$\pm$0.12 &1.35$\pm$0.04&8.47$\pm$0.14 &12.65$\pm$0.29 &0.08$\pm$0.01 &1.95$\pm$0.04 &5.64$\pm$0.23 &0.91$\pm$0.03 &0.18&0.32$\pm$0.01&0.06$\pm$0.01&44.87\\
${T_{10}}$&5.25$\pm$0.04 &1.36$\pm$0.01&7.62$\pm$0.17 &13.59$\pm$0.31 &0.13$\pm$0.01 &2.09$\pm$0.05 &3.30$\pm$0.12 &1.95$\pm$0.33 &0.14&0.67$\pm$0.12&0.13$\pm$0.02&44.85\\
${F_6}$&5.70$\pm$0.05 &1.44$\pm$0.01&7.60$\pm$0.12 &11.80$\pm$0.13 &0.41$\pm$0.02 &1.82$\pm$0.02 &1.52$\pm$0.05 &5.80$\pm$0.58 &0.31&2.01$\pm$0.20&0.40$\pm$0.04&44.52\\
${T_{11}}$&5.80$\pm$0.03 &1.55$\pm$0.01&7.37$\pm$0.14 &12.36$\pm$0.09 &0.46$\pm$0.03 &1.90$\pm$0.01 &1.50$\pm$0.08 &6.99$\pm$0.74 &0.16&2.42$\pm$0.26&0.48$\pm$0.05&44.55\\
${B_6}$&5.14$\pm$0.01 &1.34$\pm$0.02&7.85$\pm$0.32 &14.16$\pm$0.23 &0.10$\pm$0.01 &2.18$\pm$0.04 &3.77$\pm$0.49 &0.32$\pm$0.13 &0.32&0.11$\pm$0.04&0.02$\pm$0.01&44.24\\
${T_{12}}$&5.78$\pm$0.14 &1.33$\pm$0.01&6.90$\pm$0.06 &11.48$\pm$0.33 &0.41$\pm$0.02 &1.77$\pm$0.05 &1.21$\pm$0.02 &4.38$\pm$0.34 &0.26&1.52$\pm$0.12&0.30$\pm$0.02&44.29\\
${F_7}$&5.81$\pm$0.02 &1.57$\pm$0.01&7.10$\pm$0.21 &9.56$\pm$0.45 &0.61$\pm$0.06 &1.47$\pm$0.07 &1.76$\pm$0.01 &2.51$\pm$0.10 &0.22&0.87$\pm$0.03&0.17$\pm$0.01&44.08\\
					\hline
			\end{tabular}}\\
		\end{adjustwidth}
	\end{center} 
	\tablecomments{Note. The columns represent the following: (1) Epoch, the state of OQ 334 during different epochs within the MJD 58678-60387; (2) $s$, the spectral index; (3) $b$, the spectral curvature; (4) $\gamma_{0}$, the initial Lorentz factor; (5) $\delta$, the Doppler factor; (6) B, the magnetic field, in units of Gs; (7) $R$, the radius of the radiation zone, specified in $cm$; (8) $r_{dis}$, the location of the radiation zone, specified in $cm$; (9) $U_{ext}$, the energy density of the external photon field, in units of $erg\cdot cm^{-3}$; (10) $\chi ^{2}$, the chi-square value; (11) $L_{disk}$, the accretion disk luminosity, in units of $erg\cdot s^{-1}$; (12) \.{m}, the accretion rate; (13) $log(P_{jet})$, the jet radiation power, in units of $erg\cdot s^{-1}$.}		
	
\end{sidewaystable*}

%------------------------------------------

\begin{acknowledgments}
We thank the anonymous editor and referee for very constructive and helpful comments and suggestions.
We acknowledge the use of data, analysis tools, and services from the Open Universe platform, the ASI Space Science Data Center (SSDC), the Automated Survey of All-Sky Supernovae (ASAS), the Astrophysics Science Archive Research Center (HEASARC), the Fermi Science Tools, the Astrophysics Data System (ADS), and the National Extra-galactic Database (NED).
This work is partially supported by the National Natural Science Foundation of China (Grant Nos. 12363002 and 12163002) 
The authors would like to express their gratitude to EditSprings (https://www.editsprings.cn ) for the expert linguistic services provided.
\end{acknowledgments}

%\software{\textit{JetSet} (\citealt{2009A&A...501..879T,2011ApJ...739...66T,2020ascl.soft09001T}), iminuit (V2.22.0) \citep{dembinski_2023_8070217}, Astropy \citep{2013A&A...558A..33A,2018AJ....156..123A,2022ApJ...935..167A}, Matplotlib \citep{2007CSE.....9...90H}, Numpy \citep{2020Natur.585..357H}.
%}

\bibliography{manuscript}{}

\begin{thebibliography}{}
\expandafter\ifx\csname natexlab\endcsname\relax\def\natexlab#1{#1}\fi
\providecommand{\url}[1]{\href{#1}{#1}}
\providecommand{\dodoi}[1]{doi:~\href{http://doi.org/#1}{\nolinkurl{#1}}}
\providecommand{\doeprint}[1]{\href{http://ascl.net/#1}{\nolinkurl{http://ascl.net/#1}}}
\providecommand{\doarXiv}[1]{\href{https://arxiv.org/abs/#1}{\nolinkurl{https://arxiv.org/abs/#1}}}

\bibitem[{{Abdo} {et~al.}(2010{\natexlab{a}}){Abdo}, {Ackermann}, {Agudo}, {Ajello}, {Aller}, {Aller}, {Angelakis}, {Arkharov}, {Axelsson}, {Bach}, {Baldini}, {Ballet}, {Barbiellini}, {Bastieri}, {Baughman}, {Bechtol}, {Bellazzini}, {Benitez}, {Berdyugin}, {Berenji}, {Blandford}, {Bloom}, {Boettcher}, {Bonamente}, {Borgland}, {Bregeon}, {Brez}, {Brigida}, {Bruel}, {Burnett}, {Burrows}, {Buson}, {Caliandro}, {Calzoletti}, {Cameron}, {Capalbi}, {Caraveo}, {Carosati}, {Casandjian}, {Cavazzuti}, {Cecchi}, {{\c{C}}elik}, {Charles}, {Chaty}, {Chekhtman}, {Chen}, {Chiang}, {Chincarini}, {Ciprini}, {Claus}, {Cohen-Tanugi}, {Colafrancesco}, {Cominsky}, {Conrad}, {Costamante}, {Cutini}, {D'ammando}, {Deitrick}, {D'Elia}, {Dermer}, {de Angelis}, {de Palma}, {Digel}, {Donnarumma}, {Silva}, {Drell}, {Dubois}, {Dultzin}, {Dumora}, {Falcone}, {Farnier}, {Favuzzi}, {Fegan}, {Focke}, {Forn{\'e}}, {Fortin}, {Frailis}, {Fuhrmann}, {Fukazawa}, {Funk}, {Fusco}, {G{\'o}mez}, {Gargano}, {Gasparrini}, {Gehrels}, {Germani},
  {Giebels}, {Giglietto}, {Giommi}, {Giordano}, {Giuliani}, {Glanzman}, {Godfrey}, {Grenier}, {Gronwall}, {Grove}, {Guillemot}, {Guiriec}, {Gurwell}, {Hadasch}, {Hanabata}, {Harding}, {Hayashida}, {Hays}, {Healey}, {Heidt}, {Hiriart}, {Horan}, {Hoversten}, {Hughes}, {Itoh}, {Jackson}, {J{\'o}hannesson}, {Johnson}, {Johnson}, {Jorstad}, {Kadler}, {Kamae}, {Katagiri}, {Kataoka}, {Kawai}, {Kennea}, {Kerr}, {Kimeridze}, {Kn{\"o}dlseder}, {Kocian}, {Kopatskaya}, {Koptelova}, {Konstantinova}, {Kovalev}, {Kovalev}, {Kurtanidze}, {Kuss}, {Lande}, {Larionov}, {Latronico}, {Leto}, {Lindfors}, {Longo}, {Loparco}, {Lott}, {Lovellette}, {Lubrano}, {Madejski}, {Makeev}, {Marchegiani}, {Marscher}, {Marshall}, {Max-Moerbeck}, {Mazziotta}, {McConville}, {McEnery}, {Meurer}, {Michelson}, {Mitthumsiri}, {Mizuno}, {Moiseev}, {Monte}, {Monzani}, {Morselli}, {Moskalenko}, {Murgia}, {Nestoras}, {Nilsson}, {Nizhelsky}, {Nolan}, {Norris}, {Nuss}, {Ohsugi}, {Ojha}, {Omodei}, {Orlando}, {Ormes}, {Osborne}, {Ozaki}, {Pacciani},
  {Padovani}, {Pagani}, {Page}, {Paneque}, {Panetta}, {Parent}, {Pasanen}, {Pavlidou}, {Pelassa}, {Pepe}, {Perri}, {Pesce-Rollins}, {Piranomonte}, {Piron}, {Pittori}, {Porter}, {Puccetti}, {Rahoui}, {Rain{\`o}}, {Raiteri}, {Rando}, {Razzano}, {Reimer}, {Reimer}, {Reposeur}, {Richards}, {Ritz}, {Rochester}, {Rodriguez}, {Romani}, {Ros}, {Roth}, {Roustazadeh}, {Ryde}, {Sadrozinski}, {Sadun}, {Sanchez}, {Sander}, {Saz Parkinson}, {Scargle}, {Sellerholm}, {Sgr{\`o}}, {Shaw}, {Sigua}, {Siskind}, {Smith}, {Smith}, {Spandre}, {Spinelli}, {Starck}, {Stevenson}, {Stratta}, {Strickman}, {Suson}, {Tajima}, {Takahashi}, {Takahashi}, {Takalo}, {Tanaka}, {Thayer}, {Thayer}, {Thompson}, {Tibaldo}, {Torres}, {Tosti}, {Tramacere}, {Uchiyama}, {Usher}, {Vasileiou}, {Verrecchia}, {Vilchez}, {Villata}, {Vitale}, {Waite}, {Wang}, {Winer}, {Wood}, {Ylinen}, {Zensus}, {Zhekanis}, \& {Ziegler}}]{2010ApJ...716...30A}
{Abdo}, A.~A., {Ackermann}, M., {Agudo}, I., {et~al.} 2010{\natexlab{a}}, \apj, 716, 30, \dodoi{10.1088/0004-637X/716/1/30}

\bibitem[{{Abdo} {et~al.}(2010{\natexlab{b}}){Abdo}, {Ackermann}, {Ajello}, {Atwood}, {Axelsson}, {Baldini}, {Ballet}, {Barbiellini}, {Bastieri}, {Bechtol}, {Bellazzini}, {Berenji}, {Blandford}, {Bloom}, {Bonamente}, {Borgland}, {Bouvier}, {Bregeon}, {Brez}, {Brigida}, {Bruel}, {Burnett}, {Buson}, {Caliandro}, {Cameron}, {Caraveo}, {Carrigan}, {Casandjian}, {Cavazzuti}, {Cecchi}, {{\c{C}}elik}, {Charles}, {Chekhtman}, {Cheung}, {Chiang}, {Ciprini}, {Claus}, {Cohen-Tanugi}, {Conrad}, {Cutini}, {Dermer}, {de Angelis}, {de Palma}, {Digel}, {Silva}, {Drell}, {Dubois}, {Dumora}, {Farnier}, {Favuzzi}, {Fegan}, {Focke}, {Fortin}, {Frailis}, {Fukazawa}, {Funk}, {Fusco}, {Gargano}, {Gasparrini}, {Gehrels}, {Germani}, {Giebels}, {Giglietto}, {Giommi}, {Giordano}, {Glanzman}, {Godfrey}, {Grenier}, {Grondin}, {Grove}, {Guillemot}, {Guiriec}, {Harding}, {Hartman}, {Hayashida}, {Hays}, {Healey}, {Horan}, {Hughes}, {Jackson}, {J{\'o}hannesson}, {Johnson}, {Johnson}, {Kamae}, {Katagiri}, {Kataoka}, {Kawai}, {Kerr},
  {Kn{\"o}dlseder}, {Kuss}, {Lande}, {Latronico}, {Lemoine-Goumard}, {Longo}, {Loparco}, {Lott}, {Lovellette}, {Lubrano}, {Madejski}, {Makeev}, {Mazziotta}, {McConville}, {McEnery}, {Meurer}, {Michelson}, {Mitthumsiri}, {Mizuno}, {Moiseev}, {Monte}, {Monzani}, {Morselli}, {Moskalenko}, {Murgia}, {Nolan}, {Norris}, {Nuss}, {Ohsugi}, {Omodei}, {Orlando}, {Ormes}, {Paneque}, {Panetta}, {Parent}, {Pelassa}, {Pepe}, {Persic}, {Pesce-Rollins}, {Piron}, {Porter}, {Rain{\`o}}, {Rando}, {Razzano}, {Reimer}, {Reimer}, {Reposeur}, {Ritz}, {Rochester}, {Rodriguez}, {Romani}, {Roth}, {Ryde}, {Sadrozinski}, {Sanchez}, {Sander}, {Saz Parkinson}, {Scargle}, {Sgr{\`o}}, {Siskind}, {Smith}, {Smith}, {Spandre}, {Spinelli}, {Strickman}, {Suson}, {Tajima}, {Takahashi}, {Takahashi}, {Tanaka}, {Thayer}, {Thayer}, {Thompson}, {Tibaldo}, {Torres}, {Tosti}, {Tramacere}, {Uchiyama}, {Usher}, {Vasileiou}, {Vilchez}, {Villata}, {Vitale}, {Waite}, {Wang}, {Winer}, {Wood}, {Ylinen}, \& {Ziegler}}]{2010ApJ...710.1271A}
{Abdo}, A.~A., {Ackermann}, M., {Ajello}, M., {et~al.} 2010{\natexlab{b}}, \apj, 710, 1271, \dodoi{10.1088/0004-637X/710/2/1271}

\bibitem[{{Abdollahi} {et~al.}(2020){Abdollahi}, {Acero}, {Ackermann}, {Ajello}, {Atwood}, {Axelsson}, {Baldini}, {Ballet}, {Barbiellini}, {Bastieri}, {Becerra Gonzalez}, {Bellazzini}, {Berretta}, {Bissaldi}, {Blandford}, {Bloom}, {Bonino}, {Bottacini}, {Brandt}, {Bregeon}, {Bruel}, {Buehler}, {Burnett}, {Buson}, {Cameron}, {Caputo}, {Caraveo}, {Casandjian}, {Castro}, {Cavazzuti}, {Charles}, {Chaty}, {Chen}, {Cheung}, {Chiaro}, {Ciprini}, {Cohen-Tanugi}, {Cominsky}, {Coronado-Bl{\'a}zquez}, {Costantin}, {Cuoco}, {Cutini}, {D'Ammando}, {DeKlotz}, {de la Torre Luque}, {de Palma}, {Desai}, {Digel}, {Di Lalla}, {Di Mauro}, {Di Venere}, {Dom{\'\i}nguez}, {Dumora}, {Fana Dirirsa}, {Fegan}, {Ferrara}, {Franckowiak}, {Fukazawa}, {Funk}, {Fusco}, {Gargano}, {Gasparrini}, {Giglietto}, {Giommi}, {Giordano}, {Giroletti}, {Glanzman}, {Green}, {Grenier}, {Griffin}, {Grondin}, {Grove}, {Guiriec}, {Harding}, {Hayashi}, {Hays}, {Hewitt}, {Horan}, {J{\'o}hannesson}, {Johnson}, {Kamae}, {Kerr}, {Kocevski}, {Kovac'evic'},
  {Kuss}, {Landriu}, {Larsson}, {Latronico}, {Lemoine-Goumard}, {Li}, {Liodakis}, {Longo}, {Loparco}, {Lott}, {Lovellette}, {Lubrano}, {Madejski}, {Maldera}, {Malyshev}, {Manfreda}, {Marchesini}, {Marcotulli}, {Mart{\'\i}-Devesa}, {Martin}, {Massaro}, {Mazziotta}, {McEnery}, {Mereu}, {Meyer}, {Michelson}, {Mirabal}, {Mizuno}, {Monzani}, {Morselli}, {Moskalenko}, {Negro}, {Nuss}, {Ojha}, {Omodei}, {Orienti}, {Orlando}, {Ormes}, {Palatiello}, {Paliya}, {Paneque}, {Pei}, {Pe{\~n}a-Herazo}, {Perkins}, {Persic}, {Pesce-Rollins}, {Petrosian}, {Petrov}, {Piron}, {Poon}, {Porter}, {Principe}, {Rain{\`o}}, {Rando}, {Razzano}, {Razzaque}, {Reimer}, {Reimer}, {Remy}, {Reposeur}, {Romani}, {Saz Parkinson}, {Schinzel}, {Serini}, {Sgr{\`o}}, {Siskind}, {Smith}, {Spandre}, {Spinelli}, {Strong}, {Suson}, {Tajima}, {Takahashi}, {Tak}, {Thayer}, {Thompson}, {Tibaldo}, {Torres}, {Torresi}, {Valverde}, {Van Klaveren}, {van Zyl}, {Wood}, {Yassine}, \& {Zaharijas}}]{2020ApJS..247...33A}
{Abdollahi}, S., {Acero}, F., {Ackermann}, M., {et~al.} 2020, \apjs, 247, 33, \dodoi{10.3847/1538-4365/ab6bcb}

\bibitem[{{Acharyya} {et~al.}(2023){Acharyya}, {Adams}, {Archer}, {Bangale}, {Benbow}, {Brill}, {Christiansen}, {Chromey}, {Errando}, {Falcone}, {Feng}, {Finley}, {Foote}, {Fortson}, {Furniss}, {Gallagher}, {Hanlon}, {Hanna}, {Hervet}, {Hinrichs}, {Hoang}, {Holder}, {Jin}, {Johnson}, {Kaaret}, {Kertzman}, {Kieda}, {Kleiner}, {Korzoun}, {Krennrich}, {Lang}, {Lundy}, {Maier}, {McGrath}, {Millard}, {Millis}, {Mooney}, {Moriarty}, {Mukherjee}, {O'Brien}, {Ong}, {Pohl}, {Pueschel}, {Quinn}, {Ragan}, {Reynolds}, {Ribeiro}, {Roache}, {Sadeh}, {Sadun}, {Saha}, {Santander}, {Sembroski}, {Shang}, {Splettstoesser}, {Talluri}, {Tucci}, {Vassiliev}, {Williams}, {Wong}, {VERITAS Collaboration}, {Hovatta}, {Jorstad}, {Kiehlmann}, {L{\"a}hteenm{\"a}ki}, {Liodakis}, {Marscher}, {Max-Moerbeck}, {Readhead}, {Reeves}, {Smith}, \& {Tornikoski}}]{2023ApJ...950..152A}
{Acharyya}, A., {Adams}, C.~B., {Archer}, A., {et~al.} 2023, \apj, 950, 152, \dodoi{10.3847/1538-4357/acd2d0}

\bibitem[{{Bednarek}(1998)}]{1998MNRAS.294..439B}
{Bednarek}, W. 1998, \mnras, 294, 439, \dodoi{10.1046/j.1365-8711.1998.01183.x10.1111/j.1365-8711.1998.01183.x}

\bibitem[{{Blandford} \& {Levinson}(1995)}]{1995ApJ...441...79B}
{Blandford}, R.~D., \& {Levinson}, A. 1995, \apj, 441, 79, \dodoi{10.1086/175338}

\bibitem[{{Cash}(1979)}]{1979ApJ...228..939C}
{Cash}, W. 1979, \apj, 228, 939, \dodoi{10.1086/156922}

\bibitem[{{Cavaliere} \& {D'Elia}(2002)}]{2002ApJ...571..226C}
{Cavaliere}, A., \& {D'Elia}, V. 2002, \apj, 571, 226, \dodoi{10.1086/339778}

\bibitem[{{Chen}(2018)}]{2018ApJS..235...39C}
{Chen}, L. 2018, \apjs, 235, 39, \dodoi{10.3847/1538-4365/aab8fb}

\bibitem[{{Dermer} {et~al.}(1992){Dermer}, {Schlickeiser}, \& {Mastichiadis}}]{1992A&A...256L..27D}
{Dermer}, C.~D., {Schlickeiser}, R., \& {Mastichiadis}, A. 1992, \aap, 256, L27

\bibitem[{{Gammie} {et~al.}(1999){Gammie}, {Narayan}, \& {Blandford}}]{1999ApJ...516..177G}
{Gammie}, C.~F., {Narayan}, R., \& {Blandford}, R. 1999, \apj, 516, 177, \dodoi{10.1086/307089}

\bibitem[{{Ghisellini} \& {Madau}(1996)}]{1996MNRAS.280...67G}
{Ghisellini}, G., \& {Madau}, P. 1996, \mnras, 280, 67, \dodoi{10.1093/mnras/280.1.67}

\bibitem[{{Ghisellini} {et~al.}(2009){Ghisellini}, {Maraschi}, \& {Tavecchio}}]{2009MNRAS.396L.105G}
{Ghisellini}, G., {Maraschi}, L., \& {Tavecchio}, F. 2009, \mnras, 396, L105, \dodoi{10.1111/j.1745-3933.2009.00673.x}

\bibitem[{{Ghisellini} {et~al.}(2013){Ghisellini}, {Tavecchio}, {Foschini}, {Bonnoli}, \& {Tagliaferri}}]{2013MNRAS.432L..66G}
{Ghisellini}, G., {Tavecchio}, F., {Foschini}, L., {Bonnoli}, G., \& {Tagliaferri}, G. 2013, \mnras, 432, L66, \dodoi{10.1093/mnrasl/slt041}

\bibitem[{{Ghisellini} {et~al.}(2011){Ghisellini}, {Tavecchio}, {Foschini}, \& {Ghirlanda}}]{2011MNRAS.414.2674G}
{Ghisellini}, G., {Tavecchio}, F., {Foschini}, L., \& {Ghirlanda}, G. 2011, \mnras, 414, 2674, \dodoi{10.1111/j.1365-2966.2011.18578.x}

\bibitem[{{Ghisellini} {et~al.}(2010){Ghisellini}, {Tavecchio}, {Foschini}, {Ghirlanda}, {Maraschi}, \& {Celotti}}]{2010MNRAS.402..497G}
{Ghisellini}, G., {Tavecchio}, F., {Foschini}, L., {et~al.} 2010, \mnras, 402, 497, \dodoi{10.1111/j.1365-2966.2009.15898.x}

\bibitem[{{Kang} {et~al.}(2023){Kang}, {Zheng}, \& {Wu}}]{2023MNRAS.525.3201K}
{Kang}, S.-J., {Zheng}, Y.-G., \& {Wu}, Q. 2023, \mnras, 525, 3201, \dodoi{10.1093/mnras/stad2456}

\bibitem[{{Kochanek} {et~al.}(2017){Kochanek}, {Shappee}, {Stanek}, {Holoien}, {Thompson}, {Prieto}, {Dong}, {Shields}, {Will}, {Britt}, {Perzanowski}, \& {Pojma{\'n}ski}}]{2017PASP..129j4502K}
{Kochanek}, C.~S., {Shappee}, B.~J., {Stanek}, K.~Z., {et~al.} 2017, \pasp, 129, 104502, \dodoi{10.1088/1538-3873/aa80d9}

\bibitem[{{Lasota} {et~al.}(1996){Lasota}, {Abramowicz}, {Chen}, {Krolik}, {Narayan}, \& {Yi}}]{1996ApJ...462..142L}
{Lasota}, J.~P., {Abramowicz}, M.~A., {Chen}, X., {et~al.} 1996, \apj, 462, 142, \dodoi{10.1086/177137}

\bibitem[{{Lyu} {et~al.}(2022){Lyu}, {Wu}, {Yan}, {Yu}, \& {Liu}}]{2022ApJ...927..227L}
{Lyu}, B., {Wu}, Q., {Yan}, Z., {Yu}, W., \& {Liu}, H. 2022, \apj, 927, 227, \dodoi{10.3847/1538-4357/ac5256}

\bibitem[{{Maraschi} {et~al.}(1992){Maraschi}, {Ghisellini}, \& {Celotti}}]{1992ApJ...397L...5M}
{Maraschi}, L., {Ghisellini}, G., \& {Celotti}, A. 1992, \apjl, 397, L5, \dodoi{10.1086/186531}

\bibitem[{{Marscher} \& {Gear}(1985)}]{1985ApJ...298..114M}
{Marscher}, A.~P., \& {Gear}, W.~K. 1985, \apj, 298, 114, \dodoi{10.1086/163592}

\bibitem[{{Mishra} {et~al.}(2021){Mishra}, {Dai}, {Chen}, {Cheng}, {Jayasinghe}, {Tucker}, {Vallely}, {Bersier}, {Bose}, {Do}, {Dong}, {Holoien}, {Huber}, {Kochanek}, {Liang}, {Payne}, {Prieto}, {Shappee}, {Stanek}, {Bhatiani}, {Cox}, {DeFrancesco}, {Shen}, {Thompson}, \& {Wang}}]{2021ApJ...913..146M}
{Mishra}, H.~D., {Dai}, X., {Chen}, P., {et~al.} 2021, \apj, 913, 146, \dodoi{10.3847/1538-4357/abf63d}

\bibitem[{{Narayan} \& {Yi}(1995)}]{1995ApJ...444..231N}
{Narayan}, R., \& {Yi}, I. 1995, \apj, 444, 231, \dodoi{10.1086/175599}

\bibitem[{{Ren} {et~al.}(2024){Ren}, {Zhou}, {Zheng}, {Kang}, \& {Wu}}]{2024A&A...685A.140R}
{Ren}, S.~S., {Zhou}, R.~X., {Zheng}, Y.~G., {Kang}, S.~J., \& {Wu}, Q. 2024, \aap, 685, A140, \dodoi{10.1051/0004-6361/202347312}

\bibitem[{{Riess} {et~al.}(2022){Riess}, {Yuan}, {Macri}, {Scolnic}, {Brout}, {Casertano}, {Jones}, {Murakami}, {Anand}, {Breuval}, {Brink}, {Filippenko}, {Hoffmann}, {Jha}, {D'arcy Kenworthy}, {Mackenty}, {Stahl}, \& {Zheng}}]{2022ApJ...934L...7R}
{Riess}, A.~G., {Yuan}, W., {Macri}, L.~M., {et~al.} 2022, \apjl, 934, L7, \dodoi{10.3847/2041-8213/ac5c5b}

\bibitem[{{Ruan} {et~al.}(2014){Ruan}, {Anderson}, {Plotkin}, {Brandt}, {Burnett}, {Myers}, \& {Schneider}}]{2014ApJ...797...19R}
{Ruan}, J.~J., {Anderson}, S.~F., {Plotkin}, R.~M., {et~al.} 2014, \apj, 797, 19, \dodoi{10.1088/0004-637X/797/1/19}

\bibitem[{{Shappee} {et~al.}(2014){Shappee}, {Prieto}, {Grupe}, {Kochanek}, {Stanek}, {De Rosa}, {Mathur}, {Zu}, {Peterson}, {Pogge}, {Komossa}, {Im}, {Jencson}, {Holoien}, {Basu}, {Beacom}, {Szczygie{\l}}, {Brimacombe}, {Adams}, {Campillay}, {Choi}, {Contreras}, {Dietrich}, {Dubberley}, {Elphick}, {Foale}, {Giustini}, {Gonzalez}, {Hawkins}, {Howell}, {Hsiao}, {Koss}, {Leighly}, {Morrell}, {Mudd}, {Mullins}, {Nugent}, {Parrent}, {Phillips}, {Pojmanski}, {Rosing}, {Ross}, {Sand}, {Terndrup}, {Valenti}, {Walker}, \& {Yoon}}]{2014ApJ...788...48S}
{Shappee}, B.~J., {Prieto}, J.~L., {Grupe}, D., {et~al.} 2014, \apj, 788, 48, \dodoi{10.1088/0004-637X/788/1/48}

\bibitem[{{Sikora} {et~al.}(1994){Sikora}, {Begelman}, \& {Rees}}]{1994ApJ...421..153S}
{Sikora}, M., {Begelman}, M.~C., \& {Rees}, M.~J. 1994, \apj, 421, 153, \dodoi{10.1086/173633}

\bibitem[{{Stocke} {et~al.}(1991){Stocke}, {Morris}, {Gioia}, {Maccacaro}, {Schild}, {Wolter}, {Fleming}, \& {Henry}}]{1991ApJS...76..813S}
{Stocke}, J.~T., {Morris}, S.~L., {Gioia}, I.~M., {et~al.} 1991, \apjs, 76, 813, \dodoi{10.1086/191582}

\bibitem[{{Urry}(1998)}]{1998AdSpR..21...89U}
{Urry}, C.~M. 1998, Advances in Space Research, 21, 89, \dodoi{10.1016/S0273-1177(97)00619-4}

\bibitem[{{Urry} \& {Padovani}(1995)}]{1995PASP..107..803U}
{Urry}, C.~M., \& {Padovani}, P. 1995, \pasp, 107, 803, \dodoi{10.1086/133630}

\bibitem[{{Wagner} \& {Witzel}(1995)}]{1995ARA&A..33..163W}
{Wagner}, S.~J., \& {Witzel}, A. 1995, \araa, 33, 163, \dodoi{10.1146/annurev.aa.33.090195.001115}

\bibitem[{{Zhou} {et~al.}(2024){Zhou}, {Zheng}, {Zhu}, {Kang}, \& {Li}}]{2024ApJ...962...22Z}
{Zhou}, R.~X., {Zheng}, Y.~G., {Zhu}, K.~R., {Kang}, S.~J., \& {Li}, X.~P. 2024, \apj, 962, 22, \dodoi{10.3847/1538-4357/ad0a66}

\end{thebibliography}
\bibliographystyle{aasjournal}

\end{document}